\documentclass[fleqn,usenatbib]{mnras}

\usepackage{newtxtext,newtxmath}
\usepackage{color}
\definecolor{magen}{rgb}{0.70,0.10,0.50}
\definecolor{dgreen}{rgb}{0.1,0.5,0.2}

\usepackage[T1]{fontenc}
\usepackage{ae,aecompl}
\usepackage{ulem}
\usepackage{hyperref}

\usepackage{graphicx}	
\usepackage{amsmath,bm}	
\usepackage{amssymb}	

\renewcommand{\Vec}[1]{\bmath{#1}}

\title[{\sc Grale} lens inversion with up to 1000 multiple images]
{{Free-form {\sc Grale} lens inversion of galaxy clusters with up to 1000 multiple images}}

\author[A. Ghosh et al.]{Agniva Ghosh$^{1}$\thanks{E-mail: ghosh116@umn.edu (AG); llrw@umn.edu(LLRW)}, Liliya L.R. Williams$^{1}$ and Jori Liesenborgs$^{2}$\\
$^{1}$School of Physics and Astronomy, University of Minnesota, 116 Church Street SE, Minneapolis, MN 55455, USA\\
$^{2}$UHasselt -- tUL, Expertisecentrum voor Digitale Media, Wetenschapspark 2, B-3590, Diepenbeek, Belgium
}

\date{Accepted XXX. Received YYY; in original form ZZZ}

\pubyear{2019}

\begin{document}
\label{firstpage}
\pagerange{\pageref{firstpage}--\pageref{lastpage}}
\maketitle

\begin{abstract}
In the near future, ultra deep observations of galaxy clusters with HST or JWST will uncover $300-1000$ lensed multiple images, increasing the current count per cluster by up to an order of magnitude. This will further refine our view of clusters, leading to a more accurate and precise mapping of the total and dark matter distribution in clusters, and enabling a better understanding of background galaxy population and their luminosity functions. However, to effectively use that many images as input to lens inversion will require a re-evaluation of, and possibly upgrades to the existing methods. In this paper we scrutinize the performance of the free-form lens inversion method \textsc{Grale} in the regime of $150-1000$ input images, using synthetic massive galaxy clusters.
Our results show that with an increasing number of input images, \textsc{Grale} produces improved reconstructed mass distributions, with the fraction of the lens plane recovered at better than $10\%$ accuracy increasing from $40-50\%$ for $\sim\!\!150$ images to $65\%$ for $\sim\!1000$ images. The reconstructed time delays imply a more precise measurement of $H_0$, with $\lesssim 1\%$ bias.
While the fidelity of the reconstruction improves with the increasing number of multiple images used as model constraints, $\sim\!\!150$ to $\sim\!\!1000$, the lens plane rms deteriorates from $\sim\!0.11\arcsec$ to $\sim\!0.28\arcsec$.  Since lens plane rms is not necessarily the
best indicator of the quality of the mass reconstructions, looking for an alternative indicator is warranted.

\end{abstract} 




\section{Introduction}

The phenomena of gravitational lensing or the bending of light in the presence of gravitational bodies provide astrophysicists with a unique opportunity to study the mass distribution in massive clusters of galaxies and to use them as powerful natural telescopes. Galaxy clusters host dark matter, gas, and stars, largely confined to galaxies. Clusters form late, are extremely massive, but gravitationally bound. Much has been learned about the distribution of dark matter and hot gas in clusters, but many questions remain, for example, relating to the fraction of dark matter in compact substructures of $\sim 10-10^8\,M_\odot$ \citep{Dai2018,diego18,oguri18,venu17}. Merging clusters are especially interesting because the nature of interactions between member galaxies and cluster dark matter depends, in part, on non-gravitational forces, and hence properties of dark matter particles \citep{Harvey2015,harvey17,massey18}. To utilize clusters of galaxies as cosmic telescopes, one needs to characterize their uneven optics, i.e. obtain magnification maps. These are derived from the mass distribution maps, making them the most important factor in the success of any gravitational lens reconstruction.

In 2005, Abell 1689 became the first cluster to be studied using $\sim$100 images \citep{Broadhurst2005}, an impressive improvement on the 10-30 images per cluster that was common previously.  Recently, in an ambitious effort the {\it Hubble Frontier Field Survey} (HFF; PI: J. Lotz) project has used six massive merging clusters at intermediate redshifts as nature's most powerful telescopes. Deep HFF observations have revealed $\sim 100-150$ multiple images per cluster. To generate mass models of the clusters, HFF has taken advantage of a range of existing lens inversion codes; about eight in all: 
\textsc{Lenstool} \citep{Jullo2009, Johnson2014, jauzac15}, 
SWUnited \citep{Bradac2009, Strait2018}, 
WSLAP+ \citep{Sendra2014, Diego2016}, 
\textsc{Grale} \citep{sebesta2019}, 
\textsc{Glafic} \citep{Kawamata2016}, 
LensPerfect \citep{Coe2010}, 
LTM \citep{Zitrin2016}, 
PIEMDeNFW \citep{Zitrin2013} etc. 
The ongoing Hubble Space Telescope's \textit{Beyond Ultra-deep Frontier Fields And Legacy Observations} (BUFFALO; PI: C. Steinhardt) project will also utilize several of these lens inversion techniques. 
Comprehensive code comparison exercises were conducted based on HFF data, using synthetic lenses \citep{Meneghetti2017}, as well as observed HFF clusters \citep{priewe2017,Gonzalez2018}, which shed light on the systematic uncertainties in the mass reconstructions, and in the recovered luminosity function parameters of distant galaxies \citep{Bouwens2017,Atek2018}. 

Lens inversion methods fall into two main categories: parametric and free-form. Parametric models are the most widely used lens inversion methods. Observed galaxies and cluster-wide dark matter halos are assigned simple forms, for example, Pseudo Isothermal Elliptical Mass Distribution \citep{Kovner1993}, non-singular elliptical isothermal density profiles,  Navarro-Frenk-White profiles \citep{Navarro1996}, or smoothed out galaxy light distribution \citep{Jullo2007,Zitrin2009}. The observed galaxy luminosity (or stellar mass or velocity dispersion) is related to the mass distribution using scaling relations, like Faber-Jackson \citep{Faber1976}. These mass components make up the basis set. Though details differ from method to method, in parametric models cluster mass distribution is always closely tied to the light distribution. In the early days of lens modeling, when image constraints were sparse, this simplification of the modeling problem was very helpful. With an increased number of images, simplifying assumptions will become a hindrance, especially if galaxies deviate from simple scaling relations, and cluster-wide mass distribution is perturbed by the merging process. Parametric methods also tend to significantly underestimate their uncertainties \citep{rodney2015,priewe2017}.

In response to the shortcoming of the parametric models, free-form methods were developed. Free-form methods like \textsc{Grale} \citep{Liesenborgs2006a,Liesenborgs2007,Mohammed2014,Meneghetti2017} are completely agnostic about light distribution, and their basis set consists of very many parameters, which collectively represent complex cluster mass distributions. Hybrid methods, like WSLAP+ combine features of free-form and parametric methods \citep{Sendra2014, Diego2016}. \textsc{Grale} uses its basis functions as building blocks, so a given galaxy or dark matter halo is represented by a number of these functions. This makes methods like \textsc{Grale} and WSLAP+ flexible. If the number of observed multiple images is small, flexibility can become a drawback. In the case of \textsc{Grale}, which constructs a large number of mass models, flexibility can lead to unastrophysical models to be included in the ensemble average maps. 

\textsc{Grale} is ideally suited for reconstructions with plentiful strong lensing data because it uses no parametric assumptions, and the number of its model parameters exceeds the number of observational constraints, allowing wider exploration of degenerate mass distributions. In combination with HST data, \textsc{Grale} was used to derive several important results. Based on the analysis of HFF clusters MACS J0416 and Abell 2744, it has been shown in an assumption-free way that total mass clusters stronger with brighter than fainter galaxies, confirming the standard biasing, and that there are no galaxy mass-light offsets larger than $10-15$ kpc \citep{sebesta2016,sebesta2019}. In MACS J0717  \citep{williams2018} and Abell 1689 \citep{Mohammed2014}, \textsc{Grale} deduced the presence of new background high redshift line of sight structures. In MACS J1149, \textsc{Grale} postdicted the position of SN Ia Refsdal image SX to $\la 0.036''$, better than most other methods \citep{williams2019}. Using Abell 2744 and MACS J0416, it was shown that \textsc{Grale} uncertainties span the range of non-overlapping errorbars derived by other methods \citep{rodney2015, priewe2017}, and are thus closest to systematic uncertainties. Thus, \textsc{Grale} brings unique strengths to the problem of cluster lens reconstruction.

Given the success of HFF, prospects for deeper Hubble Space Telescope, and/or James Webb Space Telescope cluster lens project(s) are bright. Future observations will have a larger number of multiple images per cluster than are currently available. They will uncover numerous very high redshift galaxies, including those in the epoch of reionization  \citep{bouwens17, Atek2018, Cowley2018, yue18, Yung2019,bhat19}, 
as well as many highly magnified individual stars in background galaxies \citep{Kelly2018, Chen2019, kaurov19}, and will provide more accurate and precise cluster mass maps, leading to a better understanding of cluster environment and dynamics, and tighter constraints on self-interacting dark matter \citep{Kahlhoefer2015, Robertson2017}. Images near cluster centers will help resolve the issue of density cores vs. cusps in clusters \citep{Limousin2016}. Ultra diffuse galaxies, currently observed mostly through their light \citep{Janssens2017,Lee2017,Janssens2019}, may also become `visible' in the reconstructed mass, either directly, or through the mass-galaxy correlation function \citep{sebesta2016}.

The main objective of this paper is to comprehensively examine the efficiency of the \textsc{Grale} algorithm in the scenarios of future observations, when a larger number of lensed images will be available as input. To do this, two synthetic clusters were created by us, generated as a superposition of parametric lensing potential forms. We also generated sources covering a range of redshifts and forward lensed them through the synthetic clusters to generate mock images. With these mock clusters and images, we performed several reconstruction exercises with \textsc{Grale}, using $\sim$150, $\sim500$, and $\sim$1000 input images. As the main figure of merit for these reconstructions, we used the fractional difference between the projected mass density of the true and \textsc{Grale}-reconstructed maps. In addition to this, we have computed the lens plane rms, as a measure of the closeness of reconstructed images and their true counterparts, and used reconstructed time delays as an estimation of the precision of Hubble's constant measurement.

In Section~\ref{sec:irtysh} we provide the details about how the synthetic lenses were created. A description of the \textsc{Grale} algorithm is given in Section~\ref{sec:grale}. Results of our reconstructions are discussed in Sections~\ref{sec:massmap}-\ref{sec:sources}, where we present several different metrics that quantify the quality of the reconstructions, as well as correlations between them.

Throughout this paper, the $\Lambda$CDM model of cosmology: flat, matter density, $\Omega_m = 0.3$, cosmological constant density, $\Omega_{\Lambda}= 0.7$, and the dimensionless Hubble constant $h = 0.7$, was used.

\section{The Synthetic cluster: Irtysh}
\label{sec:irtysh}

\begin{table*}
    \centering
    \begin{minipage}{\textwidth}
    \caption{Parameter values of the two synthetic clusters, Irtysh I and Irtysh II. The subscripts s and b indicate the parameters related to the smaller and the bigger mass clumps. Superscripts 1 and 2 denote the first and second big mass clumps. The values of $m_*$ are normalized by $\Sigma_{\rm{crit,0}}=c^2/ 4 \pi G D_{ol}= 0.314$g/cm$^2$.}    
    \label{tab:parameters}
    \begin{center}
    \begin{tabular}{l c c c c c c c c c c c r} 
        \hline
         {\bf Model} & \#b & \#s & $\alpha_{b}$ & $\alpha_{s}$ & $b_{b}$ & $q_{b}$ & $q_{s}$ & $K_{b}^1$ & $K_{b}^2$ &$K_{s}^{all}$ & s & $m_*$\\ 
         \hline
        \textbf{Irtysh I} & 2 & 115 & 0.5 & 0.45 & 26.9 & 1.0 & 1.0 & 0.35 & -0.25 & 0 & 3.0\arcsec & 0.4\\
        \textbf{Irtysh II} & 2 & 86 & 0.6 & 0.45 & 22.5 & 1.0 & 1.0 & -0.25 & 0.35 & 0 & 3.0\arcsec & 0.5\\
        \hline
    \end{tabular}
    \end{center}
    \end{minipage}
\end{table*}

\begin{figure*}
    \includegraphics[width=0.9\textwidth]{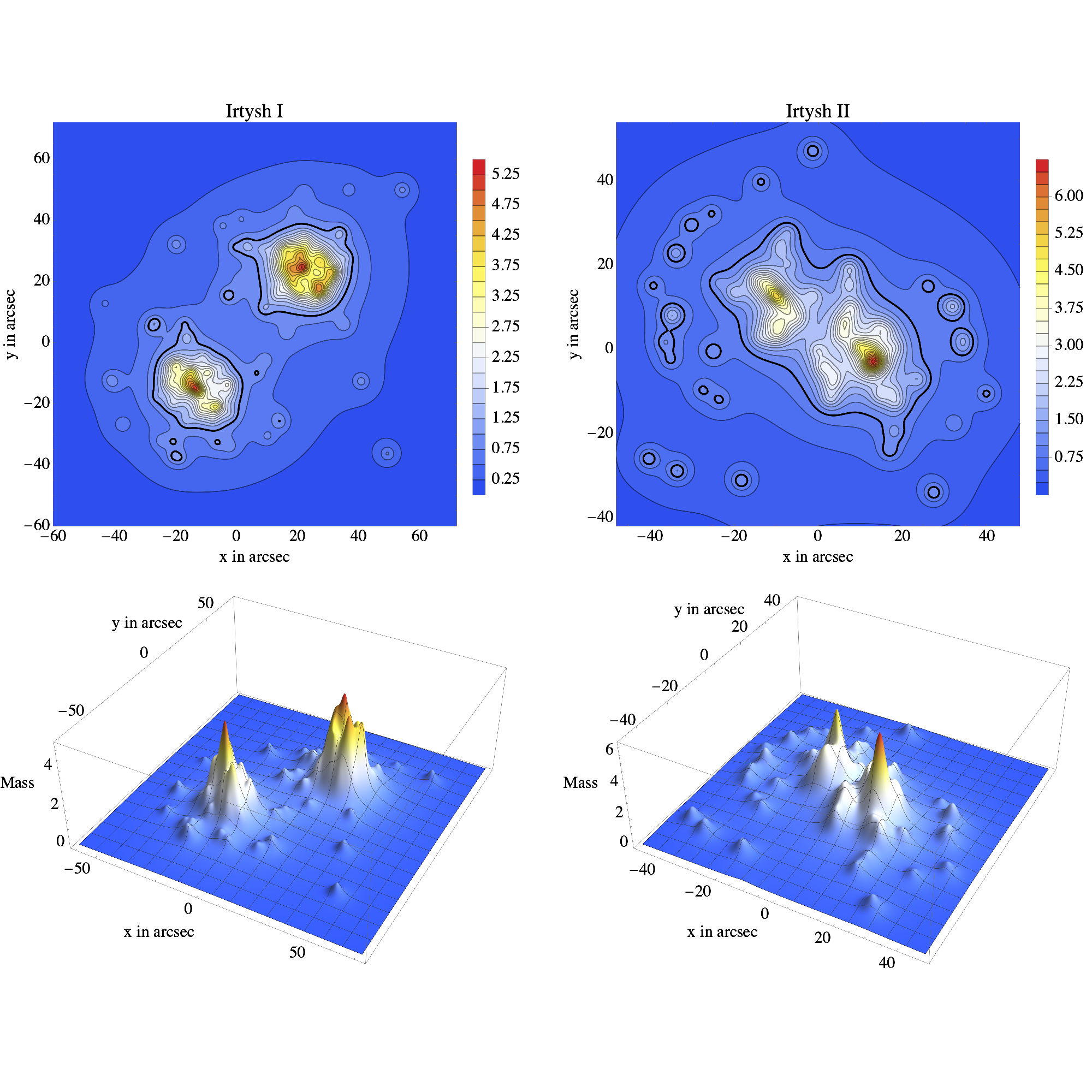}
    \caption{Projected true mass distribution of the synthetic clusters of galaxies: Irtysh I (Left Panels) and Irtysh II (Right Panels). The mass distribution is normalized by $\Sigma_{\rm{crit,0}}=c^2/ 4 \pi G D_{ol}= 0.314$g/cm$^2$.  Black contours in the top two panels show unit convergence, $\kappa=\Sigma/\Sigma_{\rm crit,0}=1$ contours.}
    \label{fig:irtysh}
\end{figure*}

\subsection{The Potential}
  We generated the mass distribution using an analytic potential. Because the cluster is synthetic, this mass distribution can be rescaled to any size. For simplicity and computational efficiency, the mass distributions for these simulated galaxy clusters were calculated using the analytic softened power-law ellipsoid potential called {\tt `alphapot'} (See Appendix~\ref{appendix:alphapot}) from the \textsc{Gravlens} catalog of models \citep{Keeton2001}, as the projected lensing potential of individual galaxies, which is of the following analytic form,
\begin{equation}
    \Psi=b(s^2+\xi^2)^{\frac{\alpha}{2}}
    \label{eq:alphapot}
\end{equation}
where $b$ is the normalization, $s$ is the core radius that eliminates the central singularity, and $\xi^2=x^2+y^2/q^2+K^2xy$ with $q$ and $K$ together representing ellipticity with non-zero position angle. Irtysh is made as a superposition of two massive cluster-scale dark matter components, and about $100$ relatively smaller galaxies. The normalizations of the smaller galaxies are determined assuming the mass function for clusters of galaxies \citep{Schechter1974,Bahcall1993},
\begin{equation}
    b_{s}=\left(\frac{m}{m_*}\right)^{-1}e^{-\frac{m}{m_*}}
    \label{eq:massfunction}
\end{equation}
where we chose $m_*$ in accordance with a more realistic visual appearance. The advantage of using an analytic potential is that the values of the deflection angles and the surface mass density can be determined exactly. We also note that the profiles we use to build up our synthetic clusters, the elliptic potential {\tt alphapot}, is different from the basis functions {\textsc{Grale}}   (see Section~\ref{sec:grale}) uses for reconstruction, which are projected Plummer spheres (see Appendix~\ref{appendix:plummers}).

\subsection{The Mass Distributions}
\label{subsection:mass}
For the analysis, we created two different mass distributions based on different parameter values in the analytic potential in Equation~\eqref{eq:alphapot}. From now on these two mass distributions will be called Irtysh I and Irtysh II. Detailed values of the parameters are given in Table~\ref{tab:parameters}. Morphology of the mass distributions are similar: both have two big mass clumps and $86$ (Irtysh I) and $115$ (II) smaller masses surrounding them. The distribution of the smaller masses is made in a way that their number decreases with the distance from the central mass peaks.  { First, the positions of the smaller lenses were generated randomly. We then used the inverse distance as probability to place them in the lens plane field. The distance was measured from the closer of the two central mass peaks. As a result, the number density of the small mass clumps decreases away from the central region.} Contour and three dimensional plots of the mass distributions are shown in Figure~\ref{fig:irtysh}.

\subsection{Sources and Images}
\label{subsection:image}

The redshift of the synthetic lenses was assumed to be $z_l=0.4$. To generate the sources, we created a number of random locations in the source plane.  We then produced a normal distribution of redshift values using the Box-Mueller algorithm, using a mean redshift of 2.8 and an upper limit of 6.5. {These values are based on existing lensed images from Frontier Fields, which already incorporate the correct source luminosity function and source redshift distributions.} Finally, these sources were forward lensed to generate the images.

\begin{table*}
    \centering
    \begin{minipage}{\textwidth}
    \caption{Details of the \textsc{Grale} reconstructions performed and the number of input images used. Details of whether a null used was also provided in the last column.}    \label{tab:runs}
    \begin{center}
    \begin{tabular}{l c c c c c c r} 
        \hline
        \textbf{Model} & \textbf{Reconstruction} & \textbf{\#Sources} & \textbf{\#Images} & \textbf{\#Minima} & \textbf{\#Maxima} & \textbf{\#Saddle} & \textbf{Null}\\
        \hline
         & \textbf{a} & 407 & 1002 & 492 & 8 & 502 & Not Used\\
        {\textbf{Irtysh I}} & \textbf{b} & 202 & 502 & 244 & 8 & 250 & Not Used\\
        {}  & \textbf{c} & 65 & 151 & 71 & 8 & 72 & Used\\
        \hline
         & \textbf{a} & 340 & 1002 & 449 & 8 & 545 & Not Used\\
        {\textbf{Irtysh II}} & \textbf{b} & 173 & 502 & 222 & 8 & 272 & Not Used\\
        {} & \textbf{c} & 53 & 151 & 64 & 8 & 79 & Used\\
        \hline
    \end{tabular}
    \end{center}
    \end{minipage}
\end{table*}

The corresponding lensed images for a given source were generated solving the lens equation, $\Vec{\theta}_I=\Vec{\beta}+\Vec{\alpha}(\Vec{\theta}_I)$. Here, $\Vec{\theta}_I$ are the image positions corresponding to a given source position $\Vec{\beta}$. The scaled deflection angle $\Vec{{\alpha}}$ was calculated using the analytic form of the first derivative of our potential {\tt alphapot}. After calculating the image positions numerically, only the sources that were multiply imaged were chosen and used for the rest of the calculations.

Once all the multiply-imaged sources and the corresponding image positions are recorded, we computed the magnification values for each image point, using analytic values of the dimensionless convergence $\kappa$ and shear $\gamma$ at the image locations. Then, based on the magnification values, the nature of the images was determined i.e. whether they are maxima, minima or saddle points in the arrival time surface. Based on this analytic classification, most of the maxima were not selected as the input images to the reconstruction process, as in reality, they are often not observable because they tend to be demagnified. Also, most of the maxima that remain are preferentially near the upper mass peak for Irtysh I and lower mass peak for Irtysh II.

\section{\textsc{Grale}}
\label{sec:grale}

\subsection{Method}
\label{subsec:fitness}

The lens inversion method used in this paper is based on the reconstruction code \textsc{Grale}. The publicly available {\textsc{Grale}} software implements a flexible, free-form, adaptive grid lens inversion method, based on a genetic algorithm \citep{Liesenborgs2006a,Liesenborgs2007,Mohammed2014,Meneghetti2017}. A {\textsc{Grale}} run starts with an initial coarse uniform grid in the lens plane populated by a basis set, such as projected Plummer density spheres (See Appendix \ref{appendix:plummers}). Each grid cell has a single Plummer sphere, with size matching the cell size. Plummer spheres are chosen because they have constant central density, and rapidly, but smoothly falling off density at larger radii. As the code runs, the denser regions are resolved with a finer grid, with each cell given a Plummer with a proportional width. The initial trial solutions, as well as all later evolved solutions are evaluated for genetic fitness, and the fit ones are cloned, combined and mutated. The final map consists of a superposition of many Plummers, typically several hundred to a couple of thousand, each with its own size and weight, determined by the genetic algorithm.

Each \textsc{Grale} run produces a slightly different final mass map, quantifying the mass uncertainties due to mass degeneracies, which exist even when the same set of image positions is used as the input. The most infamous of these degeneracies, the mass-sheet degeneracy (MSD), is broken in most of the cases due to the use of sources at multiple redshifts. Some of the other degeneracies are documented in \citet{saha2000} and \citet{Liesenborgs2012}.

\textsc{Grale} uses a combination of two types of fitness measures. (i) Positions of point-like images - a successful reconstruction would map back the lens plane images from the same source to the exact source location in the source plane. If the back projected images of the same source lie closer together, the trial solution is considered to have better fitness. The distances between the source points are not measured on an absolute scale, however. Instead, the size of the area of all back projected points of all sources is used as a length scale. This helps avoid over focusing. A better fitness for a mass map means the back projected images will have a greater fractional degree of overlap. (ii) The null space - a certain region in the image plane which has no lensed features. A trial solution may perform well with fitness (i) but still, predict unobserved additional images. To discourage \textsc{Grale} from producing these, an area larger than the region of the observed images is subdivided into a grid of triangles. By counting the number of back projected triangles overlapping the envelope of the back projected images, trial solutions are penalized if they predict additional images. \textsc{Grale} also has the capability of using critical curves, and time delays as additional fitness measures \citep{Liesenborgs2009, Mohammed2015}, but these are not utilized in the present work.

\subsection{Input}

For the two different mass distributions discussed in Section~\ref{subsection:mass}, three sets of images were prepared, consisting of about 1000, 500 and 150 images. The 150 images scenario replicates the current observational situations. The 500 and 1000 images scenarios represent future observational possibilities. 

\begin{figure}
    \includegraphics[width=\columnwidth]{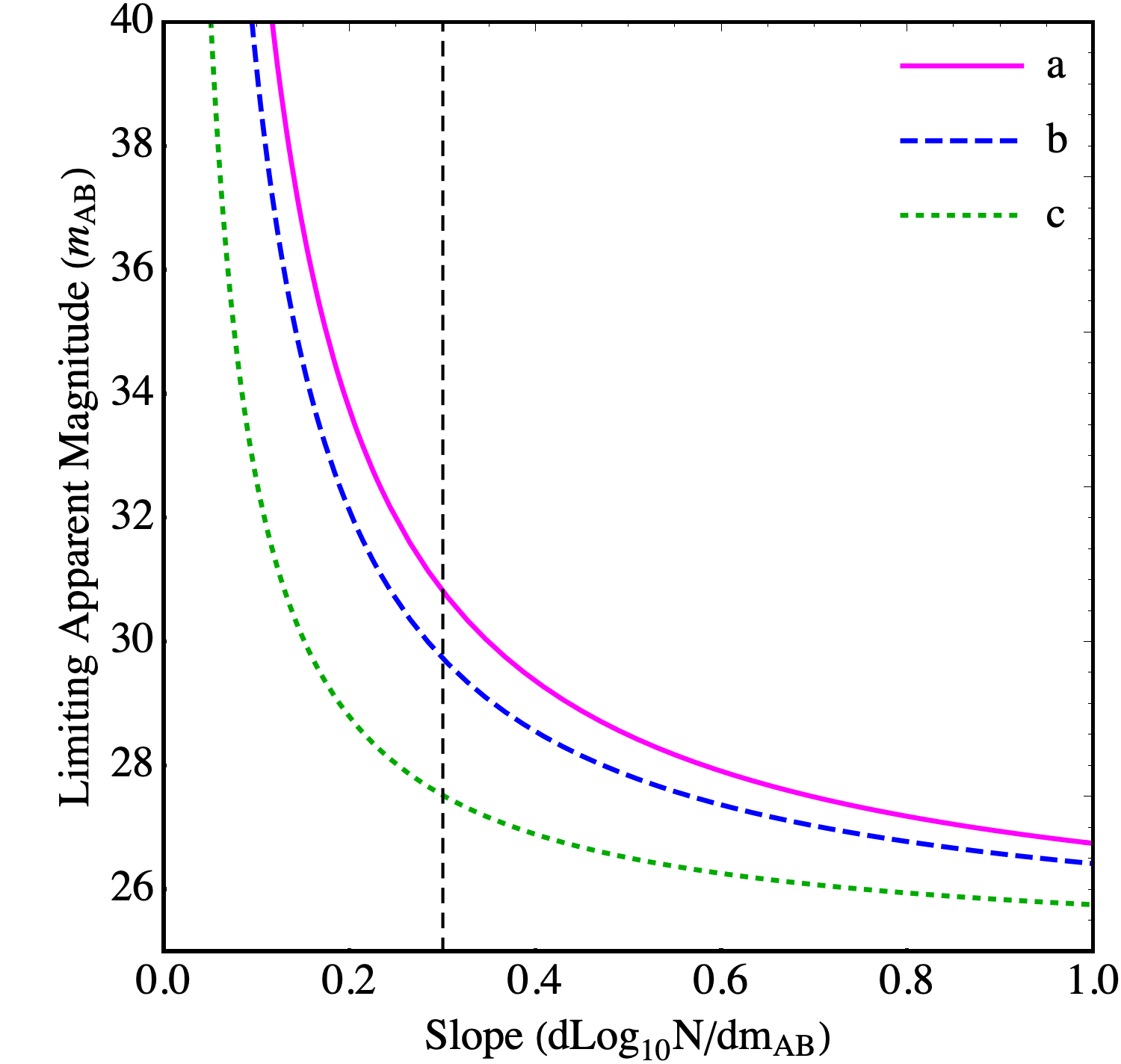}
    \caption{ Dependence of the limiting apparent magnitude ($m_{AB}$) on the slope of the power law portion of the number counts of lensed images. The solid magenta line, dashed blue line and dotted green line represent the cases for the three sets of images: a (1000 images), b (500 images) and c (150 images). For example, if the slope is 0.3 (vertical black dashed line) the limiting magnitudes would need to be 27.5, 29.7 and 30.8 for the 150, 500 and 1000 image scenarios, respectively, assuming the same lens plane area as in \citet{Jauzac2014}}
    \label{fig:mag}
\end{figure}

To translate these numbers of images into a more observationally relevant quantity, we extrapolate the existing lensed image number counts to fainter fluxes and get the rough limiting magnitudes for the three sets of reconstructions (a, b and c). For this, we used the data from \citet{Jauzac2014}, extrapolating their results with 200 images to 500 and 1000 image scenarios, based on the F814W-band AB-magnitudes provided in their Table 2. Assuming the slope of the power law portion of the number counts to be a free parameter, and $m_{\rm AB}^*=25$, we calculate the relation between the slope and the limiting magnitude, and present it in Figure~\ref{fig:mag}. As an example, the vertical dashed line corresponds to the slope value of 0.3. The limiting magnitude should be about 27.5, 29.7 and 30.8 for the 150, 500 and 1000 image scenarios, respectively, assuming the same lens plane area as in \citet{Jauzac2014}. Rescaling to a different area will change the numbers somewhat.

The input to \textsc{Grale} consists only of the image location and related redshifts. Details of the reconstructions performed with the two synthetic clusters are given in Table~\ref{tab:runs}. The 1000, 500 and 150 images cases are called reconstruction Xa, Xb, and Xc where X=(I, II) for Irtysh I and II. Each of these reconstructions is an average of 40 \textsc{Grale} runs, produced based on the set of images given in Table~\ref{tab:runs}. We did not use the null for the cases of 500 and 1000 images because the images themselves should provide sufficient constraints. Furthermore, using the null would be computationally expensive. Detailed analysis of the results obtained is discussed in the next sections.

\section{Reconstructed Mass maps}
\label{sec:massmap}

\begin{figure*}
    \includegraphics[height=630pt]{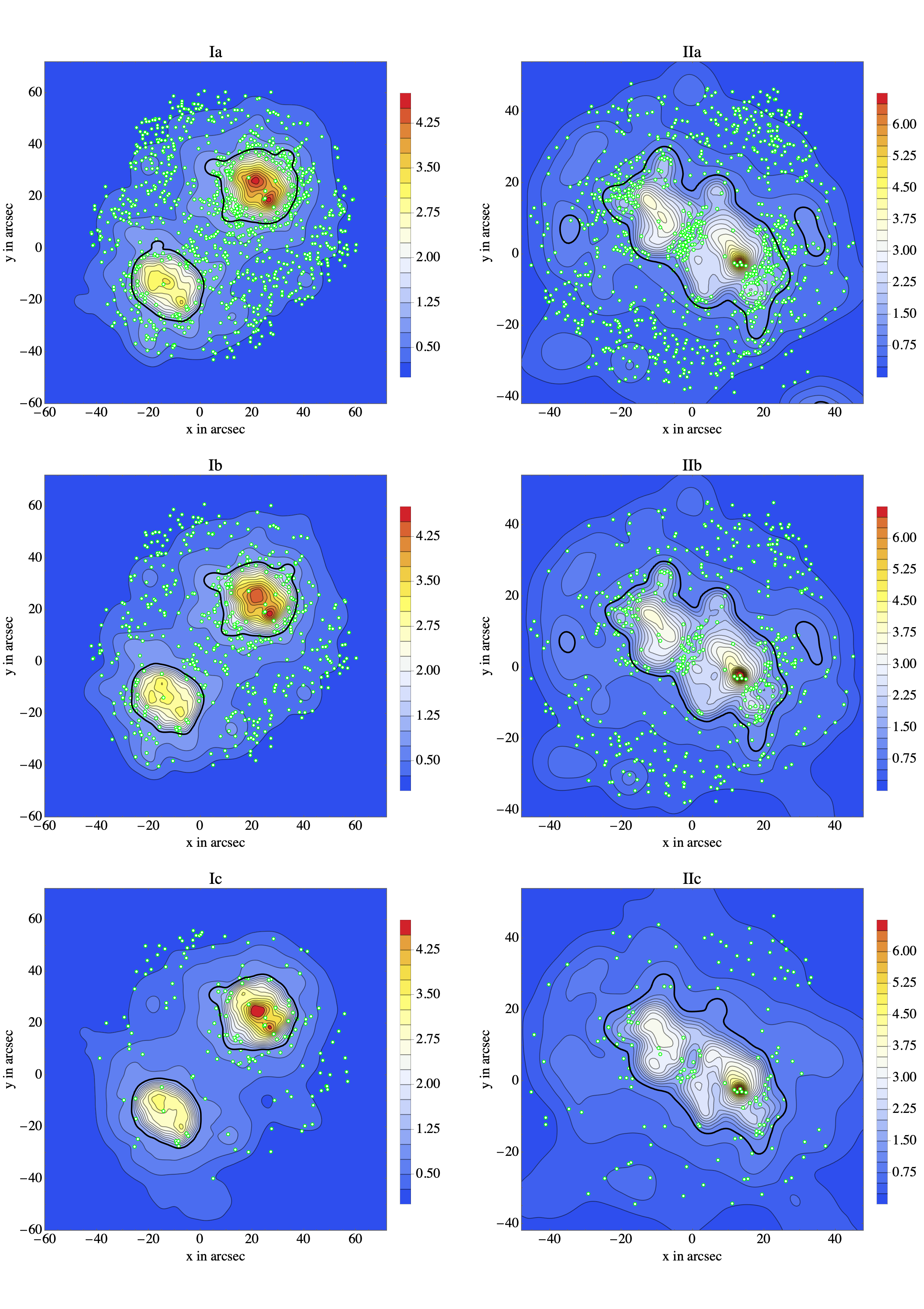}
    \caption{Projected reconstructed mass distribution of the synthetic clusters of galaxies: Irtysh I (Left Panels) and Irtysh II (Right Panels) for the different reconstructions described in Table~\ref{tab:runs}. The mass distribution is normalized by $\Sigma_{\rm{crit,0}}=c^2/ 4 \pi G D_{ol}$. Black contours in the top two panels show unit convergence, $\kappa=\Sigma/\Sigma_{\rm crit,0}=1$ contours. The green circles represents the image locations for the set of images used for each of the reconstructions.}
    \label{fig:remassmaps}
\end{figure*}

\begin{figure*}
    \includegraphics[height=630pt]{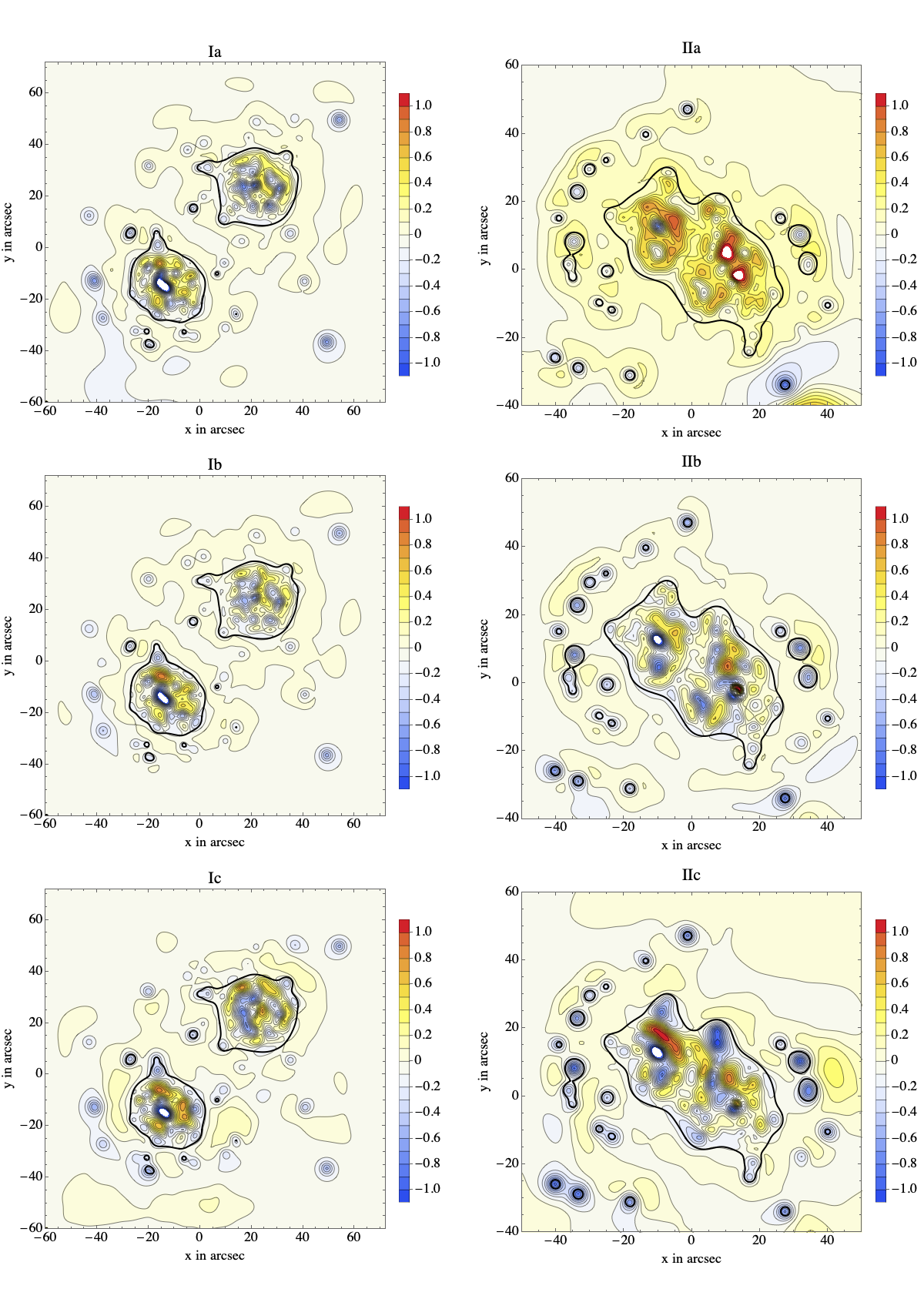}
    \caption{ Absolute mass difference maps ($\Delta m$; see Equation~\ref{eq:diff}) of the synthetic clusters of galaxies: Irtysh I (Left Panels) and Irtysh II (Right Panels) for the different reconstructions described in Table~\ref{tab:runs}. The color bar is set up in a way such that more red means the reconstructed mass is overestimated compared to the true mass, while more blue means the reconstructed mass is underestimated. The distributions are normalized by $\Sigma_{\rm{crit,0}}=c^2/ 4 \pi G D_{ol}$. Black contours show unit convergence, $\kappa=\Sigma/\Sigma_{\rm crit,0}=1$, for true mass distributions.}
    \label{fig:diffmaps}
\end{figure*}

\begin{figure*}
    \includegraphics[height=630pt]{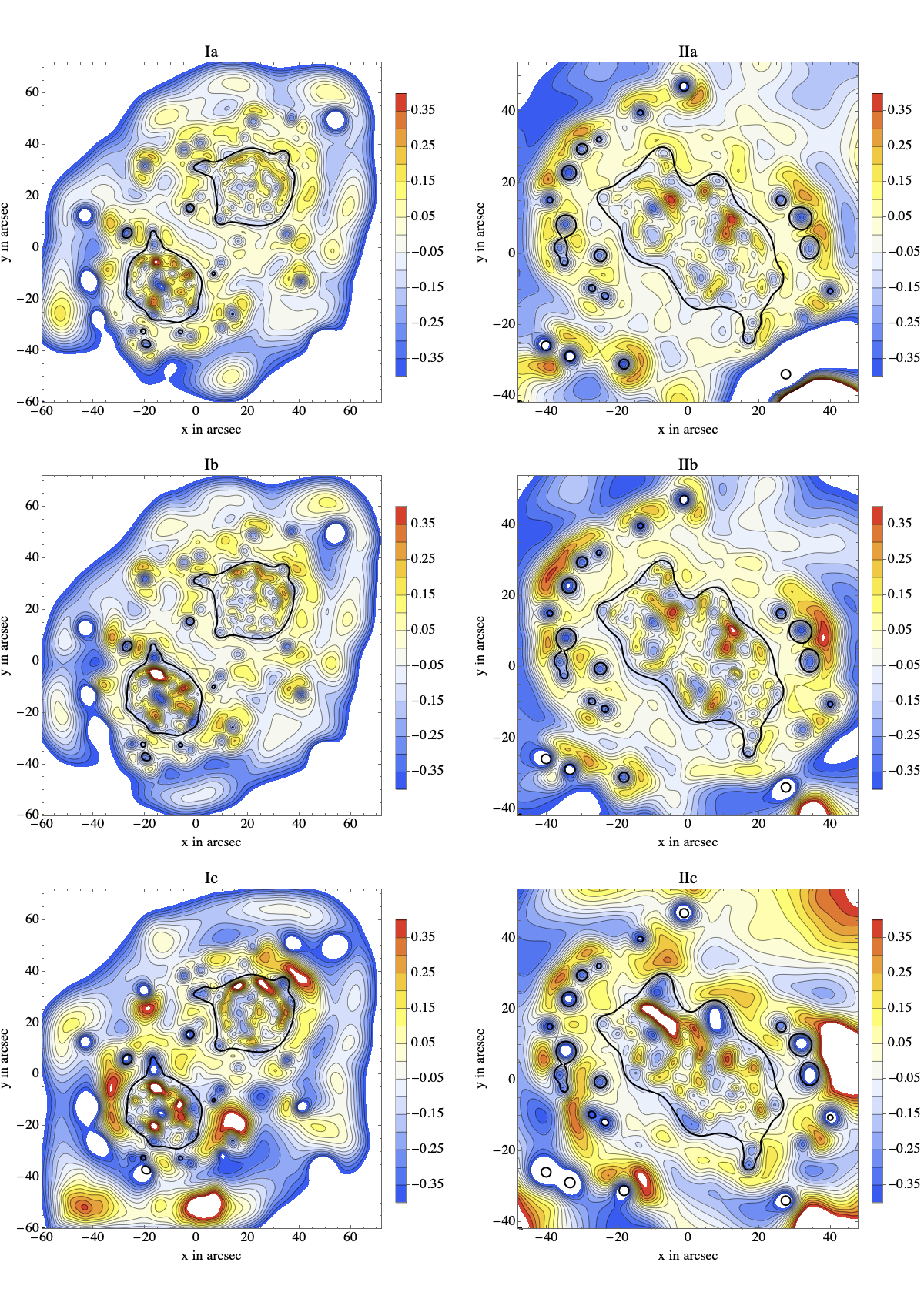}
    \caption{Fractional mass difference maps ($\Delta m/m $; see Equation~\ref{eq:fdiff}) for the 6 reconstructions. { Black contours show unit convergence, $\kappa=\Sigma/\Sigma_{\rm crit,0}=1$ for true mass distributions.} Each plot is marked for its corresponding reconstruction name on the top. The color bar is set up in a way such that more red means the reconstructed mass is overestimated compared to the true mass, while more blue means the reconstructed mass is underestimated. { White areas, for example, the lower right region of IIa, correspond to regions where mass is overestimated or underestimated and is outside the range of the color legend.}}
    \label{fig:fdiffmaps}
\end{figure*}

\begin{figure*}
    \includegraphics[height=630pt]{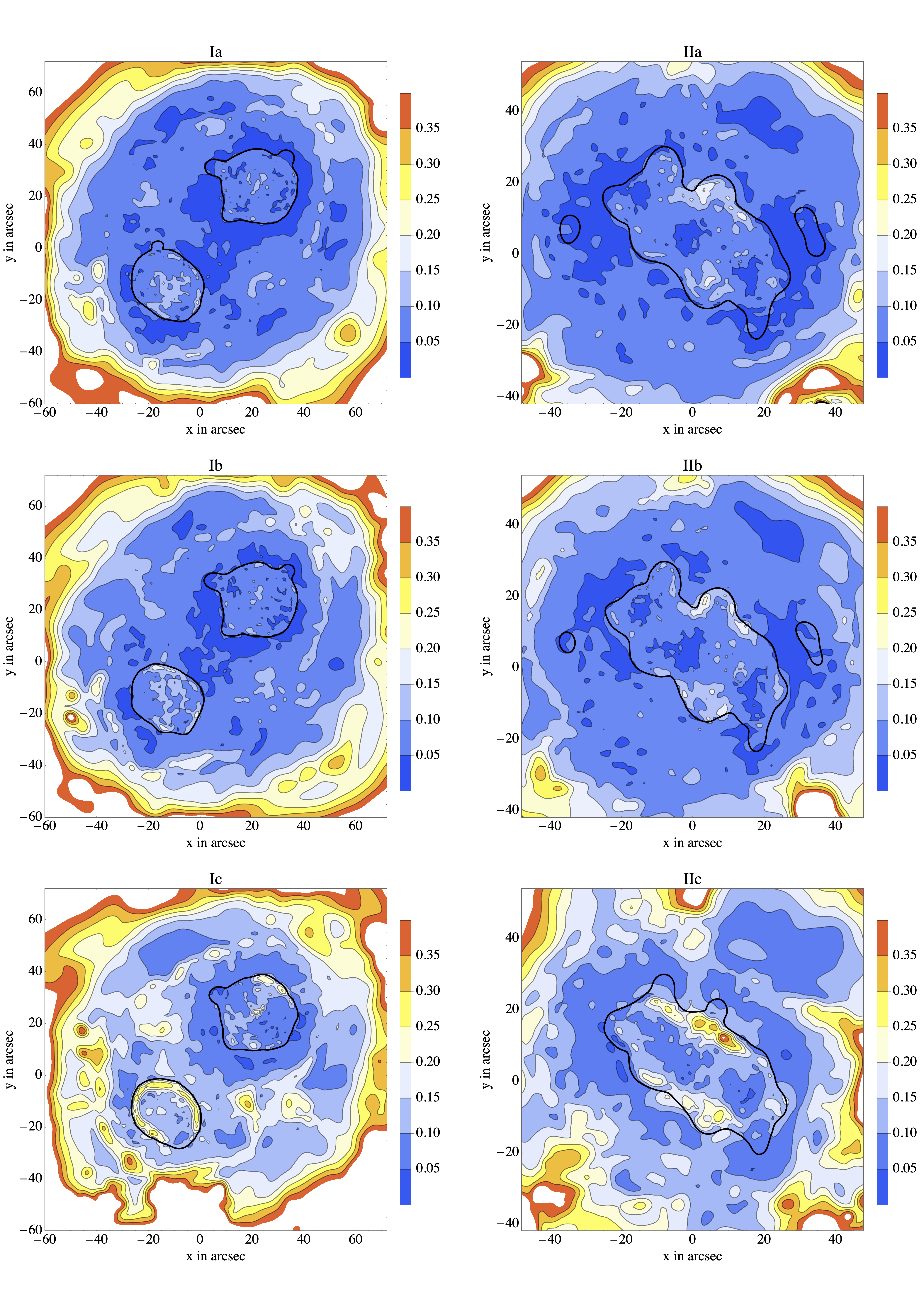}
    \caption{Fractional mass dispersion maps ($ \delta m_{\text{disp}}$; see Equation~\ref{eq:massdisp}) for the 6 reconstructions. { Black contours show unit convergence, $\kappa=\Sigma/\Sigma_{\rm crit,0}=1$ for reconstructed mass distributions.} Dispersion was calculated from 40 \textsc{Grale} runs making up each reconstruction.  Each plot is marked for its corresponding reconstruction name on the top. The color bar indicates that a deeper blue region of the lens plane exhibits narrower dispersion in reconstructed mass.}
    \label{fig:rmsmaps}
\end{figure*}

\begin{figure*}
    \includegraphics[width=0.72\textwidth]{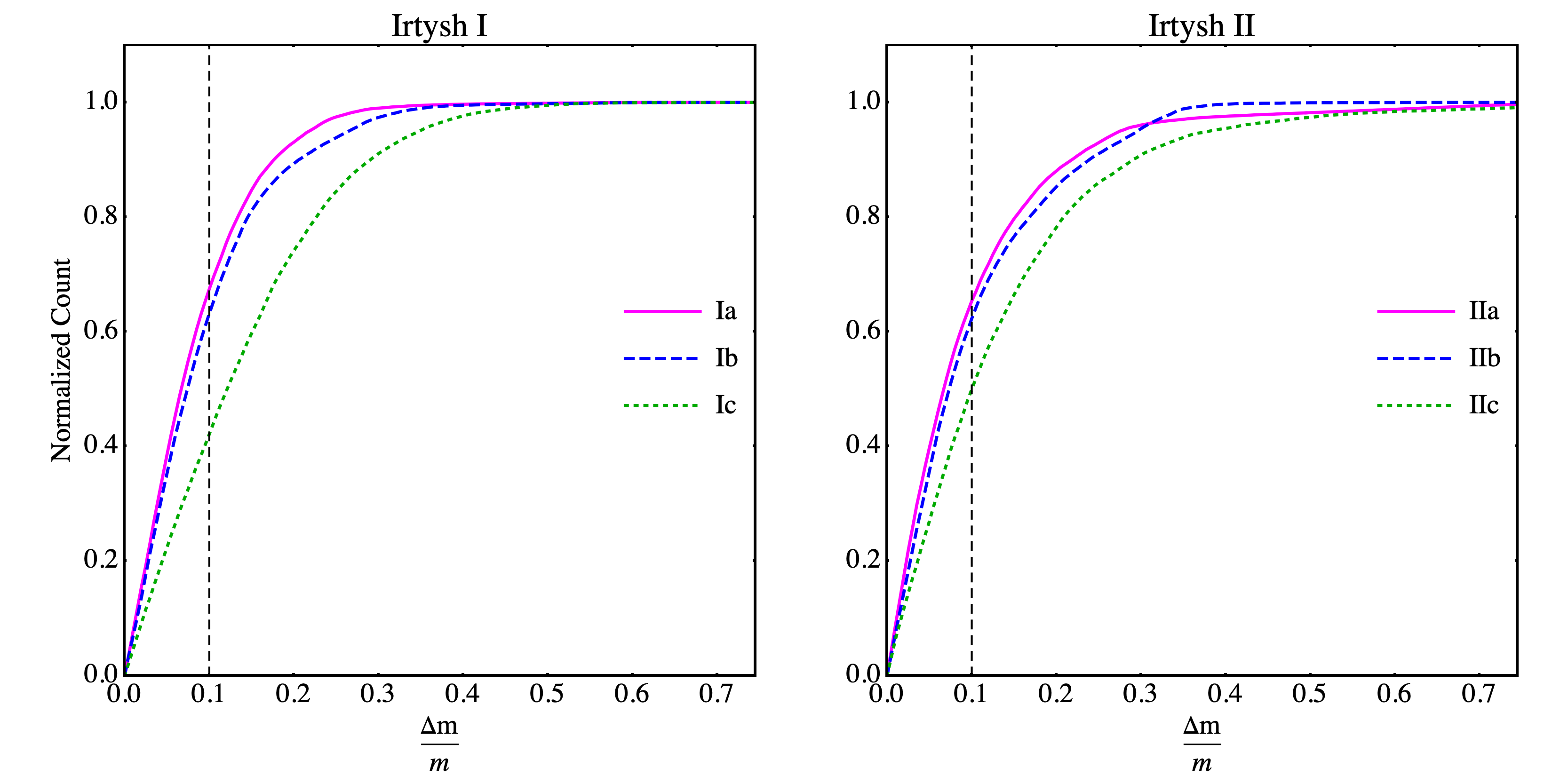}
    \caption{Cumulative histograms of fractional mass difference values {for the lens plane area containing lensed images.} The left panel presents the results of Irtysh I and the right panel represents the results of Irtysh II. For each of the panels, the solid magenta lines present the 1000 images case (reconstructions a), the 500 images scenarios (reconstructions b) are depicted by the blue dashed lines and the green dotted lines show the results from the 150 images case (reconstructions c).}
    \label{fig:chist}
\end{figure*}

\begin{figure*}
    \includegraphics[width=0.69\textwidth]{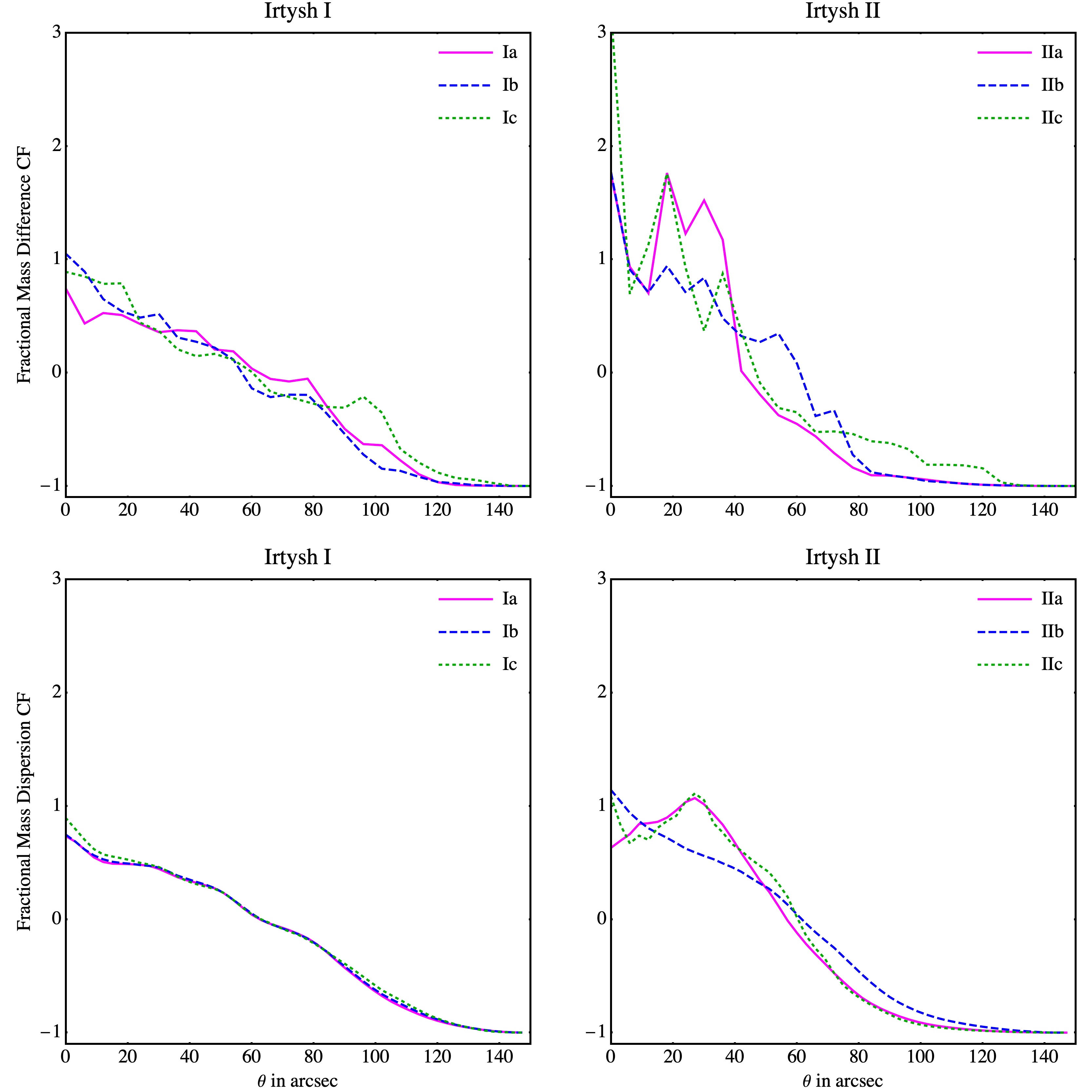}
    \caption{{Correlations that quantify the quality of reconstructions (top panels), and the uncertainties in the mass distributions (bottom panels) as a function of local surface number density of images (see Equation~\ref{eq:fdiffcorrel} and Equation~\ref{eq:rmscorrel}, respectively).} The left and right panels present the results for Irtysh I and II, respectively. In each panel, the solid magenta lines present the 1000 images case (reconstructions a), the 500 images scenarios (reconstructions b) are depicted by the blue dashed lines, and the green dotted lines show the results from the 150 images case (reconstructions c).}
    \label{fig:correlfdiff}
\end{figure*}

\begin{figure*}
    \includegraphics[width=0.9\textwidth]{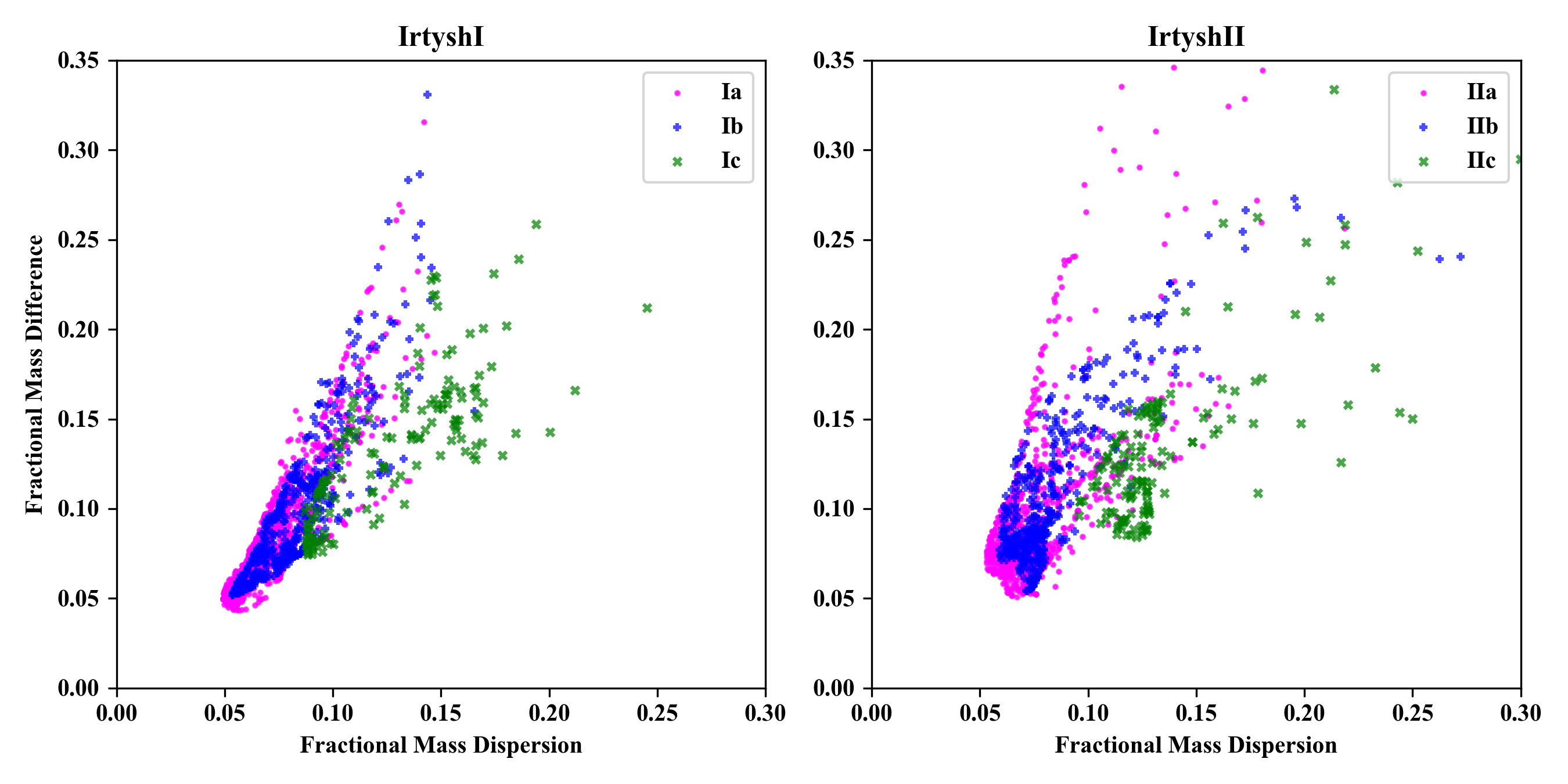}
    \caption{Fractional mass difference, $\Delta m/m $, versus fractional mass dispersion, $\delta m_{\text{disp}}$, of the lens plane pixels containing true image locations. The magenta dots, blue cross and green saltires represent the $1000$ (reconstructions a), $500$ (reconstructions b) and $150$ (reconstructions c) images scenarios, respectively. The left and right panels show results for Irtysh I and II, respectively.}
    \label{fig:fdiffrms}
\end{figure*}

Contours of the average projected mass distribution, from 40 \textsc{Grale} runs of the two synthetic clusters Irtysh I and II, are shown in Figure~\ref{fig:remassmaps}. The contours represent projected surface mass density $\Sigma$, scaled by $\Sigma_{\rm{crit,0}}=c^2/ 4 \pi G D_{ol}= 0.314$g/cm$^2$. The green circles represent the positions of the set of images used as input, for each of the reconstructions. From the reconstructed mass distributions, it is quite evident that with increasing input image numbers, more subtle substructures of the true mass distribution are recovered. For both Irtysh I and II, the central peaks are not reconstructed very well but for one of the peaks, which had fewer images near it, the reconstruction is worse compared to the other peak.   { In general having more images is always beneficial, however, the presence of central images (maxima in the arrival time surface) does not always guarantee better constrained central regions \citep{Wagner2019}. For a circularly symmetric lens the deflection angle is determined only by the total mass interior to a point, and the more central a point is, the less mass is enclosed by it. We suspect that for non-symmetric lenses the situation is similar.\footnote{  The left (or right) columns of figure 6 and 7 in \citet{Wagner2019}, shows how well one can reconstruct the central regions of synthetic clusters. In figure 7 the red circles show the image locations and the size of the circles represents the difference between reconstructed and true convergence values. For figure 6, the small blue dots represent the true convergence at the central image locations and black points represent reconstructed convergences. For every model there is at least one central image present, and there still is a difference between the true and reconstructed convergences.}}

{ The {\it absolute mass difference} maps of the reconstructed and the original mass maps are presented in Figure~\ref{fig:diffmaps}. The absolute mass difference is defined as,
\begin{equation}
    {\Delta m}=m_{\text{reconstructed}}-m_{\text{true}}.
    \label{eq:diff}
\end{equation}
These maps are also scaled by $\Sigma_{\rm{crit,0}}$, and the black contours represent unit convergence lines of the reconstructed mass distributions. The color bar is set up such that more red means the reconstructed mass is overestimated compared to the true mass, while more blue means the reconstructed mass is underestimated.}

As our main figure of merit we define the dimensionless \textit {fractional mass difference},
\begin{equation}
    \frac{\Delta m}{m}=\frac{m_{\text{reconstructed}}-m_{\text{true}}}{m_{\text{true}}}.
    \label{eq:fdiff}
\end{equation}
To calculate this quantity we divide the lens plane into small pixels{, in our post-reconstruction analysis. We call these pixels {\it lens plane pixels}.}\footnote{Lens plane pixels are used in the analysis only, to divide up the region into smaller sections and calculate the fractional difference and later other quantities. The lensing mass distributions are not pixelated.}
So, for a given region of the lens plane, the closer this quantity is to zero, the better is the reconstruction in that region. The fractional mass difference can only be computed for synthetic clusters like Irtysh, as for real clusters the true mass distribution is unknown. Figure~\ref{fig:fdiffmaps} represents the distribution of fractional mass difference values for each of the reconstructions. From the maps, it can be seen that most of the area covered by the images has a very low fractional mass difference value. 

One of the problems faced by free-form reconstructions like \textsc{Grale} is that extra mass clumps can be produced in reconstructed mass distributions because they improve the fitness values. These mass clumps are produced in the regions of the lens plane where the number of images is low, or no images are present. In lens inversions that use parametric methods, the reconstructed mass distribution in these underconstrained regions is based mostly on priors, not lensed images. In free-form lens inversions, such priors are not used, and so in these regions, mass distribution is subject to degeneracies. The resulting mass clumps can be easily noticed in the fractional mass difference maps in Figure~\ref{fig:fdiffmaps}. For reconstruction IIa, big extra mass clumps are seen towards the lower right corner, and for reconstruction IIc, towards the right side. Also for reconstruction Ic, extra mass clumps are present in the lower region, although those are smaller in size, compared to those in Irtysh II. In addition to these large mass clumps, much smaller mass excesses and deficits are present throughout the reconstructions.

{ To quantify whether using a larger set of images is producing a better overall reconstruction, we use cumulative histograms of the fractional mass difference values in the lens plane area containing all the lensed features (images). For Irtysh I, it is a circular region with a radius of $54\arcsec$ centered at $(9\arcsec, 9\arcsec)$. For Irtysh II, the circular region that has a radius of $48\arcsec$ with center at $(0\arcsec, 6\arcsec)$.} These histograms are shown in Figure~\ref{fig:chist}. The solid magenta, dashed blue and dotted green histograms represent the 1000, 500 and 150 image scenarios: Irtysh I in the left panels, and Irtysh II in the right panels. The vertical dashed black line indicates the fractional mass difference value of 0.1. From these histograms, it can be seen that for both Irtysh I and II, the 1000 image cases have $\ga 65\%$ of the pixels with better than $\Delta m/m=0.1$. In contrast to that, in the $150$ image cases, these pixels cover only $40-50\%$ of the lens plane area, whereas the $500$ images scenarios cover about $60-65\%$.

{ To determine how the quality of the reconstructions depends on the local surface number density of images, we obtained correlations defined by the following equation,
\begin{equation}
    \omega_{\Delta m/m}(\Vec{\theta}) = 1-\frac{\Big\langle N \left[\frac{1}{(\Delta m/m)^j}(\Vec{\theta}_{\rm image}^{i}\Vec{\theta}_{\rm lensplane}^{j})\right]\Big\rangle}{ \Big\langle N\left[\frac{1}{(\Delta m/m)^j}(\Vec{\theta}_{\rm random}^{i}\Vec{\theta}_{\rm lensplane}^{j})\right] \Big \rangle}
    \label{eq:fdiffcorrel}
\end{equation}
where $\Vec{\theta}_{\rm image}$ are the image locations used in the runs, $\Vec{\theta}_{\rm random}$ are the randomly generated points (same in number as the number of images), and $\Vec{\theta}_{\rm lensplane}$ are the locations of lens plane pixels. $\Vec{\theta}$ is the separation between $\Vec{\theta}_{\rm image}$ or $\Vec{\theta}_{\rm random}$ and $\Vec{\theta}_{\rm lensplane}$, and $N[\cdots]$ denotes the numbers of pairs with separation $\Vec{\theta}$, where each pair is weighted by the inverse of the fractional mass difference value $\Delta m/m$ for the corresponding lens plane pixels. These correlations} are shown in the top panel of Figure~\ref{fig:correlfdiff}. The correlation functions look similar for all three scenarios: 1000 (solid magenta) 500 (dashed blue) and 150 (dotted green) images, for both Irtysh I and II. A strong correlation can be seen between the image locations and pixels with a smaller fractional mass difference, which demonstrates that the reconstruction is better in lens plane regions near the image locations.
 
In addition to quantifying the accuracy of the reconstructions, we also quantify their precision. Each of our 6 reconstructions is an average of 40 \textsc{Grale} runs, which allows us to estimate the dimensionless \textit{fractional mass dispersion}. A lower fractional mass dispersion value indicates fewer degeneracies in the mass maps. A zero fractional mass dispersion value would imply that there are no degeneracies detected in the reconstructed mass distributions.

The fractional mass dispersion of the reconstructed mass distributions for the $i$-th pixel is defined as,
\begin{equation}
    \delta m^i_{\text{disp}}=\frac{1}{m^{i}_{\text{avg}}}\sqrt{\frac{1}{N}{\sum_{j=1,N} |m^i_j-m^i_{\text{avg}}|^2}}
    \label{eq:massdisp}
\end{equation}
where $j=1,N$, and $N=40$ is the number of \textsc{Grale} runs. The fractional mass dispersion maps are shown in Figure~\ref{fig:rmsmaps}. The reconstructions Ia and IIa have narrower dispersion compared to the Ic and IIc reconstructions. {The dependence of the uncertainty in the mass distribution on the local surface number density of images is shown in the bottom panel of Figure~\ref{fig:correlfdiff}. The correlation function in this case is defined as,
\begin{equation}
    \omega_{\delta m_{\text{disp}}}(\Vec{\theta}) = 1-\frac{\Big\langle N\left[\frac{1}{\delta m^j_{\text{disp}}}(\Vec{\theta}_{\rm image}^{i}\Vec{\theta}_{\rm lensplane}^{j})\right]\Big \rangle}{{\Big \langle}N\left[\frac{1}{\delta m^j_{\text{disp}}}(\Vec{\theta}_{\rm random}^{i}\Vec{\theta}_{\rm lensplane}^{j})\right]{\Big \rangle}}
    \label{eq:rmscorrel}
\end{equation}
which is exactly the same as Equation~\ref{eq:fdiffcorrel}, except that the weight factor is now the inverse of the fractional mass dispersion value $\delta m_{\text{disp}}$ for the corresponding lens plane pixels.
Again} a strong correlation is detected, implying that the fractional mass dispersion is smaller near the image locations, and hence lensing degeneracies are better constrained closer to images.

To obtain an additional measure on how well the regions containing images were reconstructed by \textsc{Grale}, we look at the relation between fractional mass difference (Equation~\eqref{eq:fdiff}) and the fractional mass dispersion (Equation~\eqref{eq:massdisp}) associated with the lens plane pixels containing the images; see Figure~\ref{fig:fdiffrms}. The fact that there is a clear correlation means that lens plane regions that have lower fractional mass dispersion and fewer degeneracies also have a smaller deviation from the true mass distribution. The correlation is nearly diagonal, regardless of the number of images used, though the dispersion is large at larger $\Delta m/m$ and $\delta m_{\text{disp}}$ values. Since fractional mass dispersion, $\delta m_{\text{disp}}$, can be obtained for every \textsc{Grale} reconstruction, this correlation can be useful for quantifying the fidelity of the reconstruction, if it is possible to obtain a general trend using synthetic clusters. 

{ Though not explored in this paper, we add that 
being free-form, \textsc{Grale} recovers all mass along the line of sight that affects image positions. Mass clumps that are not at the redshift of the main cluster lens are also included in the recovered mass, 'weighted' as if they were at the redshift of the cluster. This is the reason why in MACS 0717, \textsc{Grale} was able to detect the presence of a mass concentration, considerably more background than the main lens \citep{williams2018}. 
}

\section{Reconstructed Images}
\label{sec:images}

\begin{table*}
    \centering
    \caption{Summary of lens plane rms values for the six reconstructions performed. The difference in number of true and reconstructed images signifies that not for all the true images the reconstructed counterparts were found. The count of reconstructed images does not include the extra (i.e., unobserved, and probably spurious) reconstructed images produced.}    \label{tab:lprms}
    \begin{minipage}{\textwidth}
    \begin{center}
    \begin{tabular}{l c c c c c c c} 
        \hline
         {\bf Model} & & {\bf\#True } & {\bf \#Reconstructed} & {\bf LPrms} & {\bf Mean Distance} & {\bf Median Distance} & {\bf Standard Deviation}\\
         & & {\bf Images} & {\bf Images} & {${\Delta}_{rms}$} & {$\bar{\Delta}$} & {$\tilde{\Delta}$} & {$\sigma_{\Delta}$}\\
         & & & & in arcsec & in arcsec & in arcsec & in arcsec\\
         \hline
         & \textbf{a} & 1002 & 991 & 0.166876 & 0.09743 & 0.059613 & 0.135549\\
         \textbf{Irtysh I} & \textbf{b} & 502 & 499 & 0.116461 & {0.0791467} & {0.050645} & 0.0855198\\
         & \textbf{c} & 151 & 151 & 0.0553622 & 0.037264 & 0.026897 & 0.0410796\\
         \hline
         & \textbf{a} & 1002 & 981 & {0.399134} & 0.173476  & 0.083139 & 0.359646\\
         \textbf{Irtysh II} & \textbf{b} & 502 & 486 & 0.316616 & 0.28344 & 0.070229 & 0.141678\\
         & \textbf{c} & 151 & 148 & 0.170244 & 0.133921 & 0.0617075 & 0.105686\\
        \hline
    \end{tabular}
    \end{center}
    \end{minipage}
\end{table*}
\begin{figure*}
    \includegraphics[width=0.9\textwidth]{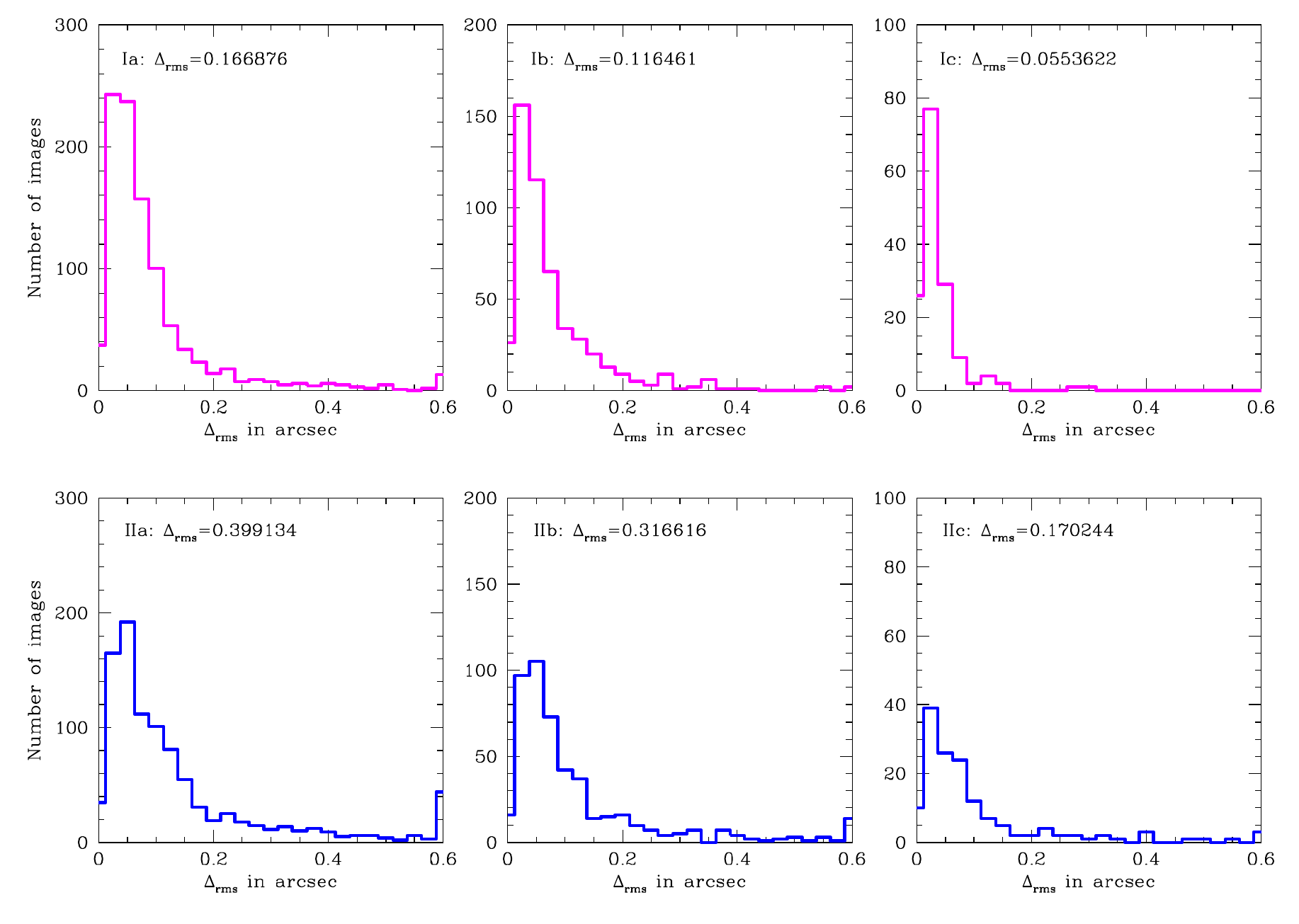}
    \caption{Histograms of the distance values of true and reconstructed images. The top panel represents the results of reconstructions of Irtysh I and the bottom panel represents the results reconstructions of Irtysh II.}
    \label{fig:lprms}
\end{figure*}

For a given surface mass density distribution and a lens inversion method, the {\it lens plane rms} quantifies how well the positions of the observed images are reproduced by a model. It can be argued that a small lens plane rms value is a necessary but not a sufficient condition to call a lens inversion good \citep{williams2018}. Lens plane rms is sometimes defined in different ways in different papers. 
The most widely used definition is,
\begin{equation}
    \Delta_{rms}^2=\frac{1}{J}\sum_{i=1,I}\left\{ \sum_{j=1,J_i} |\Vec{\theta}_{\text{true};i,j}-\Vec{\theta}_{\text{reconstructed};i,j}|^2 \right\}
    \label{eq:lprms}
\end{equation}
The main idea behind this definition is that each reconstruction uses images from $I$ number of sources, with each source having a multiplicity of $J_i$ i.e. number of images produced by that source is $J_i$, so that the total number of true input images becomes $J=\sum_i J_i$.  
Each of these true images $j$ of the source $i$, located at $\Vec{\theta}_{\text{true};i,j}$ is back projected from the lens plane to the source plane using the lens equation and deflection angle values obtained from mass distribution resulting from the lens inversion. 
Then, averaging over the $J_i$ back projected image positions gives the reconstructed source position.

These reconstructed sources can be forward lensed to obtain the reconstructed images at locations $\Vec{\theta}_{\text{reconstructed};i,j}$. At the end, the distances squared between the true images and reconstructed images are calculated as 
\begin{equation}
    \Delta^2=|\Vec{\theta}_{\text{true};i,j}-\Vec{\theta}_{\text{reconstructed};i,j}|^2
\end{equation}
These values can be directly plugged into Equation~\eqref{eq:lprms} to obtain the lens plane rms.

The main challenge in performing this task is to successfully identify the reconstructed counterparts for every true image. The most helpful ways include searching for the nearest neighbor images from a given source and determining the parity of each true image and their reconstructed counterparts. The main problem faced during this calculation concerns sources whose true and reconstructed multiplicities do not match, i.e. extra reconstructed images are produced, or for some true images, the reconstructed counterparts are not found. Extra images can be produced because \textsc{Grale} sometimes creates extra mass clumps towards the edges of the map, which results in better fitness values. These clumps can be easily seen from the fractional mass difference maps in Figure~\ref{fig:fdiffmaps}. The extra images can be eliminated based on their location and parities. Situations, where reconstructed counterparts are not produced, have to be resolved differently; we discuss these below.

Summary of lens plane rms values obtained for the six reconstructions performed are given in Table~\ref{tab:lprms}. The histograms of the lens plane rms values for each reconstruction are shown in Figure~\ref{fig:lprms}.

The third and fourth columns in Table~\ref{tab:lprms} show the numbers of true and reconstructed images; the difference between these is the number of true images for which reconstructed counterparts were not found. The true images that do not have reconstructed counterparts happen to lie near the central mass peaks, which are under-constrained.

The lens plane rms values were found to be reasonably small for all of the reconstructions, as shown in Table~\ref{tab:lprms}. HFF clusters have $\lesssim 150$ images, and lens plane rms values of $\lesssim 0.5''$ for typical reconstructions using, for example, \textsc{Lenstool} or \textsc{Grale}. For reconstructions Ic and IIc these values are smaller, $0.055''$ and $0.17''$, probably because exact image locations and redshifts were used as input for all $\sim 150$ images. 

Table~\ref{tab:lprms} clearly shows that there is an increase in lens plane rms values when the number of input images is increased. The trend is the same for both the synthetic clusters, differing only in numerical values. One can also see that the trend is strongest in the lens plane rms, weaker in the mean distance (6th column), and weakest in the median distance (7th column).
Because of the known sensitivity of rms and averages to outliers, these findings imply that outliers are partly responsible for the trends.

The fitness values of the \textsc{Grale} runs, i.e. the quantity used as the figure of merit for the solutions produced in each \textsc{Grale} run (see Section~\ref{subsec:fitness}) correlate with the lens plane rms values: the fitness values turned out to be the best for the $150$ images scenario and worse for $1000$ images.

This trend of deteriorating lens plane rms with an increasing number of images is opposite to the improvement in the reconstructed mass distributions with the increasing number of images we saw earlier (Figure~\ref{fig:fdiffmaps} and \ref{fig:chist}).  Though surprising, this behavior is not totally unexpected. With fewer input images, getting a good reconstruction is easier because of the wide range of degenerate mass solutions. As the number of images increases the range of possible degenerate mass solutions gets smaller, and as a consequence, the limitations of the reconstruction methods become more pronounced; it becomes harder to get a better fit for all the images at the same time.  This appears to be true for parametric and hybrid methods as well; for example, see figure 3 of \citet{Johnson2016} and Section 2.7 of \citet{Vega2019}. In the case of free-form methods, while the use of a large number of the basis functions allows great flexibility, the recovered model is still only an approximation of the true mass model, as is the case with parametric and hybrid methods. 

We also note that the lens plane rms values and the fitness values of Irtysh I are significantly better than those of Irtysh II. This is consistent with the result that Irtysh I is better reconstructed than Irtysh II, possibly because the latter has a more clustered image distribution than the former.
A more detailed examination of what factors are responsible for the deterioration in lens plane rms in contrast to the improvement in the mass distribution in \textsc{Grale} reconstructions, would require a further investigation.

\section{Reconstructed Time Delays}
\label{sec:timedelay}

\begin{table*}
    \centering
    \caption{Summary of the values of ratio of reconstructed time delays over the true time delays at true image location $(p^{\text{TT}})$. The differences in the numbers of true and reconstructed time delays are because of the presence of negative time delays in the reconstruction, {which were not included.}}    \label{tab:tdo}
    \begin{minipage}{\textwidth}
    \begin{center}
    \begin{tabular}{l c c c c c c c} 
        \hline
         {\bf Model} & & {\bf\#True } & {\bf \#True} & {\bf \#Reconstructed} & {\bf Mean} & {\bf Median} & {\bf Median Absolute Deviation}\\
         & & {\bf Images} & {\bf Time Delays} & {\bf Time Delays} & {$\bar{p}^{\text{TT}}$} & {$\tilde{p}^{\text{TT}}$} & MAD$(p^{\text{TT}})$ \\
         \hline
         & \textbf{a} & 1002 & 595 & 589 & {1.84923} & {0.999146} & {0.00938775}\\
         \textbf{Irtysh I} & \textbf{b} & 502 & 300 & 298 & {1.00445} & {0.997596} & {0.00788178}\\
         & \textbf{c} & 151 & 86 & 86 & {1.03079} & {1.03508} & {0.0116497}\\
         \hline
         & \textbf{a} & 1002 & 662 & 656 & {2.22199} & {1.00139} & {0.00840069}\\
         \textbf{Irtysh II} & \textbf{b} & 502 & 329 & 327 & {3.01623} & {1.01106} & {0.00916241}\\
         & \textbf{c} & 151 & 98 & 97 & {1.1148} & {1.01409} & {0.0180351}\\
        \hline
    \end{tabular}
    \end{center}
    \end{minipage}
\end{table*}
\begin{table*}
    \centering
    \caption{Summary of the values of ratio of reconstructed time delays at reconstructed image locations over the true time delays at true image location $(p^{\text{RT}})$. The differences in the numbers of true and reconstructed time delays are because of the presence of both the negative time delays { (which were not included)} in the reconstruction and the reconstructed counterpart-less true images.}    \label{tab:tdr}
    \begin{minipage}{\textwidth}
    \begin{center}
    \begin{tabular}{l c c c c c c c} 
        \hline
         {\bf Model} & & {\bf\#True } & {\bf \#True} & {\bf \#Reconstructed} & {\bf Mean} & {\bf Median} & {\bf Median Absolute Deviation}\\
         & & {\bf Images} & {\bf Time Delays} & {\bf Time Delays} & {$\bar{p}^{\text{RT}}$} & {$\tilde{p}^{\text{RT}}$} & MAD$(p^{\text{RT}})$ \\
         \hline
        & \textbf{a} & 1002 & 595 & 583 & {0.996172} & {0.999486} & {0.00789496} \\
         \textbf{Irtysh I}& \textbf{b} & 502 & 300 & 295 & {1.01716} & {0.994002} & {0.0165835}\\
         & \textbf{c} & 151 & 86 & 86 & {1.03154} & {1.03527} & {0.0112512} \\
         \hline
         & \textbf{a} & 1002 & 662 & 635 & {1.01912} & {1.00147} & {0.007146}\\
         \textbf{Irtysh II}& \textbf{b} & 502 & 329 & 314 & {1.09854} & {1.01119} & {0.00781871}\\
         & \textbf{c} & 151 & 98 & 97 & {1.057} &  {1.01483} & {0.0188137}\\
        \hline
    \end{tabular}
    \end{center}
    \end{minipage}
\end{table*}

\begin{figure*}
    \includegraphics[width=0.85\textwidth]{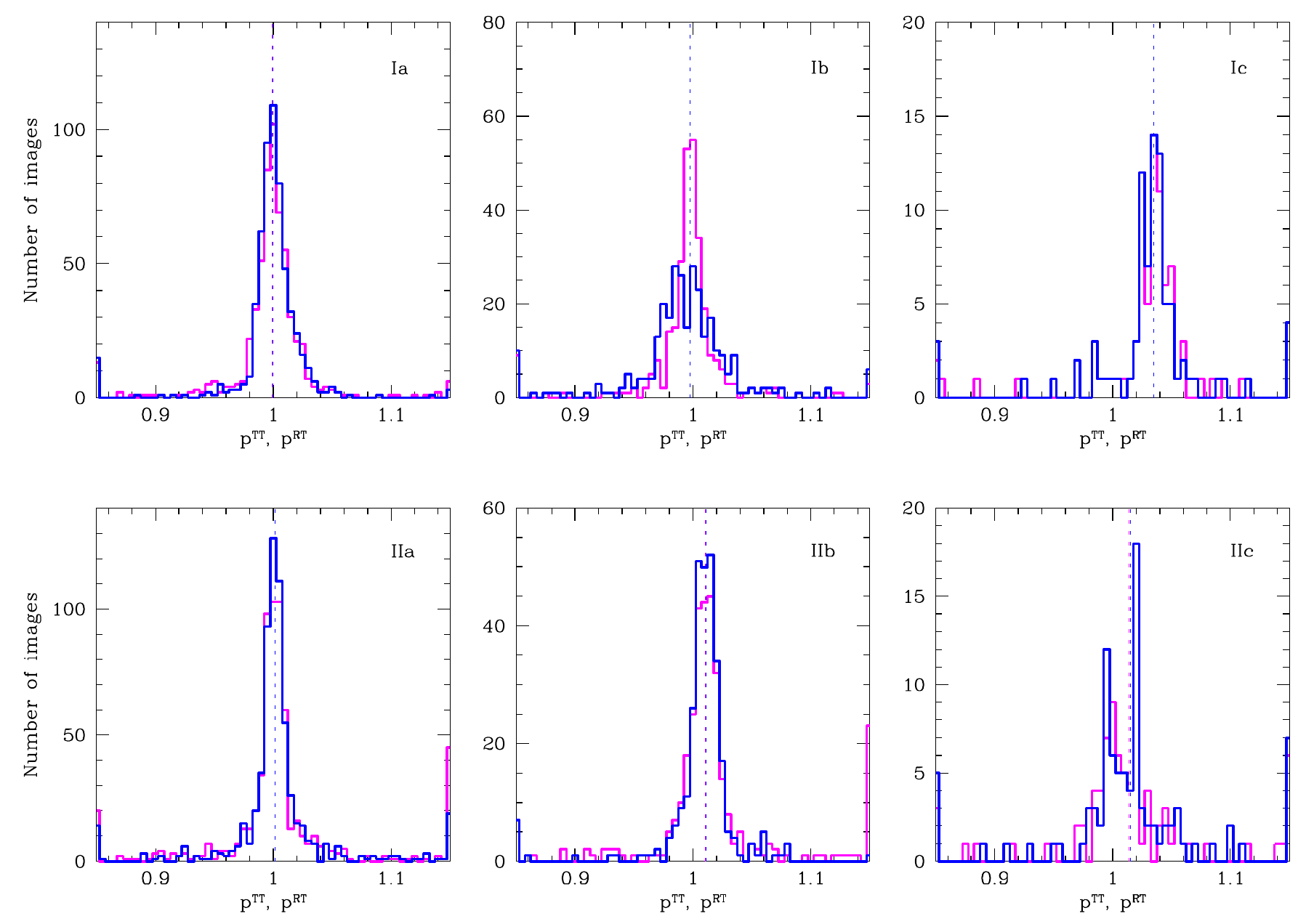}
    \caption{Histograms of the ratios of reconstructed and true time delays. The top panel represents the results of reconstructions of Irtysh I and the bottom panel represents the results reconstructions of Irtysh II. The magenta histograms represent the values of $p^{\text{TT}}$ and the blue histograms represent the values of $p^{\text{RT}}$. The vertical magenta dotted line represents the median values of $p^{\text{TT}}$ and the blue vertical dotted line represents the median values of $p^{\text{RT}}$. The medians are very close, so it is very hard to distinguish the magenta and blue vertical lines.}
    \label{fig:thist}
\end{figure*}

\begin{figure*}
    \includegraphics[width=0.75\textwidth]{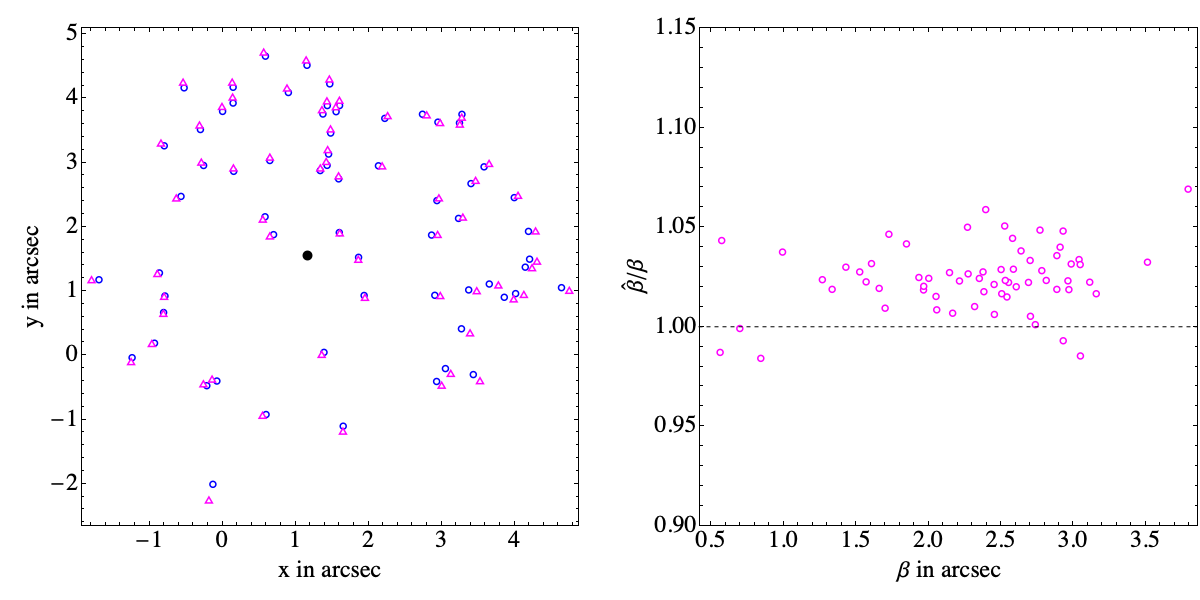}
    \caption{\textit{Left Panel:} the true (blue circles) and reconstructed (magenta triangles) source locations for reconstruction Ic. The filled black dot represents the transformation origin. \textit{Right Panel:} plot of the ratio of the reconstructed and true source positions (both w.r.t. the transformation origin) as a function of the true source positions. Note that most points form a horizontal line at $\hat\beta/\beta\approx 1.025$, which is suggestive of mass sheet degeneracy.}
    \label{fig:spt}
\end{figure*}

The time delay between the true images and the reconstructed images can be used as a one-step method to determine the Hubble's constant $H_0$ \citep{Refsdal1964}. In this section, instead of estimating the value of the Hubble constant, we perform an investigation of how much precision in $H_0$ is possible based solely on the different number of input lensed image positions provided to \textsc{Grale}. Because we are using parametric synthetic potentials to produce true images, true time delays can be calculated exactly. 
Time delays based on reconstructed mass distributions can be calculated at both the true image locations and the reconstructed image locations.  

For a given source, the ratio of reconstructed and true time delays can be presented in two ways: 
\begin{equation}
    \text{(i) }  p^{\text{TT}}=\frac{\Delta \tau_{\text{reconstructed}}  (\Vec{\theta_{\text{true}}})}{\Delta \tau_{\text{true}} (\Vec{\theta_{\text{true}}})}
\end{equation}
where $\Delta \tau_{\text{reconstructed}}(\Vec{\theta_{\text{true}}})$ are time delays based on reconstructed mass distribution at the positions of the true images, and $\Delta \tau_{\text{true}} (\Vec{\theta_{\text{true}}})$ are the true time delays, also for the true image locations (see Table~\ref{tab:tdo}).
\begin{equation}
    \text{(ii) }p^{\text{RT}}=\frac{\Delta \tau_{\text{reconstructed}}  (\Vec{\theta_{\text{reconstructed}}})}{\Delta \tau_{\text{true}} (\Vec{\theta_{\text{true}}})}
\end{equation}
where $\Delta \tau_{\text{reconstructed}}(\Vec{\theta_{\text{reconstructed}}})$ are time delays based on reconstructed mass distribution at the positions of reconstructed images 
(see Table~\ref{tab:tdr}).

Table~\ref{tab:tdo} and Table~\ref{tab:tdr} show the statistical results for the values of $p^{\text{TT}}$ and $p^{\text{RT}}$, respectively. While calculating the reconstructed  time delays, a few images produced negative values, because the arrival sequence of the recovered images is opposite compared to the true sequence. These were not included in further calculations. This creates a difference in the number of true and reconstructed time delays. In Table~\ref{tab:tdr} this difference, $<5\%$, is slightly larger than in  Table~\ref{tab:tdo}, $<1\%$, due to the presence of true images for which the reconstructed counterparts were not found.

Histograms of the time delay ratios, $p^{\text{TT}}$ and $p^{\text{RT}}$, for all of the reconstructions of both Irtysh I and Irtysh II are presented in Figure~\ref{fig:thist}. The magenta histograms are of $p^{\text{TT}}$ values, and the blue histograms represent the values of $p^{\text{RT}}$. The vertical dotted lines show the median values. 

To quantify the goodness of reconstructions in terms of time delays, we look at the median values for each of the reconstructions, in Table~\ref{tab:tdo} and Table~\ref{tab:tdr}. Averages (means) are also presented, but these are sensitive to outliers, as can be seen by comparing the mean and the median for Irtysh IIa and IIb in Table~\ref{tab:tdo}. 

The closer the median of these ratios is to unity, the better the prediction of $H_0$. For both clusters and both cases, the median improves with the increasing number of images. In a real observational situation, it is likely that only 1 or 2 time delays will be available for any given cluster. This would correspond to drawing 1 or 2 values from the histograms.
The width of the histograms, which we quantify using \textit{median absolute deviation} (MAD), gives the prediction for the precision of $H_0$, which can be achieved by the corresponding lens models.  
In general, a higher number of input images allow one to achieve higher accuracy and precision in the measurement of $H_0$, using the \textsc{Grale} reconstruction algorithm.

\section{Reconstructed Sources}
\label{sec:sources}

Lens inversion techniques are, unfortunately, plagued by various degeneracies. One kind of degeneracy faced in lens reconstruction is the \textit{source plane transformation} (SPT)  \citep{Schneider2014}, which is a generalization of the \textit{mass sheet degeneracy} (MSD). The exact MSD is broken in our case, because of the use at multiple source redshifts. But the generalized and approximate MSD can still exist, despite the presence of sources at multiple redshifts \citep{Liesenborgs2008}. The SPT can contribute to the reconstructed mass maps, and affect the reconstructed time delays, but does not affect the observed image positions. SPT changes the deflection law for all of the sources. It can be described as a transformation of the source locations in the source plane,
\begin{equation}
    \hat{\Vec{\beta}}=[1+f(\Vec{\beta})]\Vec{\beta},\label{eq:spt}
\end{equation}
where $\Vec\beta$ and $\hat{\Vec{\beta}}$ are the true and reconstructed source locations with respect to the transformation origin, and $f(\Vec{\beta})$ is a function of the true source positions. For the MSD, $[1+f(\Vec{\beta})]$ is a constant. To examine this kind of degeneracy, one needs to find the location of the transformation origin i.e. where $f(\Vec{\beta}=0)=0$, and from where the true and reconstructed source locations are measured.

To determine if SPT is the main degeneracy in the reconstructions performed in this paper, we scrutinized the positions of the true and reconstructed sources. In most reconstructions, the two sets of source positions did not exhibit any discernible pattern, from which we conclude that degeneracies more complicated than SPT are involved. The one exception was reconstruction Ic. 

The true (blue circles) and reconstructed (magenta triangles) source locations for Ic are presented in the left panel of Figure~\ref{fig:spt}. 
The transformation origin, represented by the filled black dot, was found as follows. Each true source position was connected to its reconstructed counterpart with a straight line. In a pure SPT, all such lines would intersect at a single point. In a more realistic situation, there will not exist a single intersection point, so one has to find a point that best approximates the radial nature of the Equation~\eqref{eq:spt} transformation.

In the right panel of the same figure, we show the ratio of the reconstructed and true source positions (both w.r.t. the transformation origin), as a function of the true source positions. The distribution can be approximated by a straight horizontal line, at $\hat\beta/\beta\approx 1.025$, reminiscent of MSD. In fact, the presence of this approximate MSD can also be deduced from Figure~\ref{fig:thist}. Of the 6 reconstructions, the one where the median deviates the most from unity is Ic, by an amount consistent with that obtained here. 

The distribution in this panel can also be compared to the top panel of figure 4 of \citet{Schneider2014}, which shows the same quantities, but for an analytically generated galaxy-scale SPT. Compared to the latter, our plot does not exhibit a well-defined pattern. This is probably because other degeneracies also contribute.

To our knowledge, this is the first attempt to isolate SPT in a reconstructed cluster mass distribution. A deeper exploration of the degeneracies that affect the \textsc{Grale} reconstructions will require a separate paper.

\section{Conclusions}

Synthetic cluster-lenses with $\sim 150-1000$ images presented in this paper mock the situation of near-future observations. We used these clusters to characterize the performance of the free-form lens inversion algorithm \textsc{Grale}. The summary of our results is given in the following paragraphs.

In Section~\ref{sec:massmap} we showed that using a larger number of input images improves the reconstructed mass distributions, recovering more subtle structures (Figure~\ref{fig:remassmaps}), for both the simulated clusters, Irtysh I and II. From Figure~\ref{fig:chist} it can be seen that in the $\sim1000$ image cases, $\sim65\%$ of the cluster lens plane area has a fractional mass difference between the true and reconstructed maps of $<10\%$. In the $\sim 150$ image cases, the corresponding area is only $40-50\%$. In addition, the dispersion in the mass density of the 40 \textsc{Grale} runs that make up a reconstruction is narrower if a higher number of input images is used (Figure~\ref{fig:rmsmaps}). We conclude that \textsc{Grale} produces more accurate and precise projected lens mass reconstructions when given more input images.

In Section~\ref{sec:images}, we saw that the lens plane rms, which quantifies how the reconstructed images reproduce the true ones, deteriorates when a larger number of input images is used. For $\sim 150$ and $\sim 1000$ image cases the average values are $0.11 $and $0.28$, respectively. 
This trend contradicts the expectation of achieving an improved lens plane rms as the mass reconstruction improves. 
We conclude that a lower lens plane rms cannot guarantee a better---more accurate and precise---mass reconstruction \citep{Vega2019}. As we have argued in Section~\ref{sec:images}, a large number of input images is more likely to reveal the limitations of a lens inversion method.
Because lens plane rms may not prove to be the best indicator of the quality of the mass reconstructions for either free-form or parametric methods, looking for an alternative indicator is warranted. For \textsc{Grale}, one could use the relation between the fractional mass difference and fractional mass dispersion (Figure~\ref{fig:fdiffrms}), quantified based on simulated clusters.

Section~\ref{sec:timedelay} quantifies how well $H_0$ can be measured with \textsc{Grale} reconstructions. Tables~\ref{tab:tdo} and~\ref{tab:tdr} show that with a higher number of input images, the median of the ratio of reconstructed and true time delays is getting closer to unity. For $\sim 150$ and $\sim 1000$ image cases the typical deviations from 1 are $2.5\%$ and $0.1\%$, respectively. 
However, in any given observed cluster, typically only 1 or 2 time delays can be measured. So the width of the histograms presented in Figure~\ref{fig:thist} are a better measure of the $H_0$ accuracy. We quantify this using a median absolute deviation. Typical values for $\sim 150$ and $\sim 1000$ image cases are $1.5\%$ and $0.82\%$, respectively.
This indicates that an accurate measurement of $H_0$ can be done in the near future using the \textsc{Grale} reconstructed lens models. 

Contrary to these findings in Irtysh I and II where the median predicted time delays are within about a percent of the true ones, in HFF cluster MACS J1149 \citep{williams2019} \textsc{Grale}'s average estimate is an about a factor of 2 below the true value. There are a few possible reasons for this; the most likely is that even though the total number of images is approximately the same in Irtysh and MACS J1149, their spatial and redshift distributions are not: in Irtysh I and II images are roughly evenly distributed across the lens plane and have a wide distribution in redshift, whereas in MACS J1149, 97 of the 123 images are in three tight groups, corresponding to the knots of the three images of the spiral galaxy at $z_s=1.489$. This explains why the mass distribution around SN Refsdal images SX and S1 is reconstructed with very low mass dispersion ($\sim 10\%$; figure 5 of \citet{williams2019}). In contrast to mass, time delays are affected by the whole lens plane. Since the images are so sparse over most of the lens plane in MACS J1149, time delays are not as well recovered, showing considerable dispersion, as well as a bias. Therefore, it is not just the total number of images, but their distribution in the lens plane and redshift that affects the recovery of time delays. 

Finally, in Section~\ref{sec:sources}, we found that lensing degeneracies affecting the reconstructions are more complicated than the simple source plane transformation. Further research will be required to characterize the degeneracies in more detail.

Overall, we conclude that \textsc{Grale} is well suited for cluster-lens mass reconstruction with future observations, and identify areas where further work is needed. 

\section*{Acknowledgements}

AG and LLRW acknowledge the Minnesota Supercomputing Institute (MSI) for their computational resources and support. We would like to thank the anonymous referee for their useful suggestions and comments.


\bibliographystyle{mnras}



\appendix

\section{Alphapot}
\label{appendix:alphapot}
In this paper, the mathematical backbone of the synthetic clusters is made up of the parametric potential `alphapot' \citep{Keeton2001}, whose mathematical form is given in the Equation~\eqref{eq:alphapot}. Here the first and second derivatives of the potential are listed, as they were useful for calculating several strong lensing properties of the synthetic clusters used e.g. convergence, magnification, time delay and most importantly to determine the deflection angles, for its use in the image finding algorithm.
The deflection angles in x and y are given by the first derivatives of the potential,
\begin{equation}
    \alpha_x=\frac{\partial \Psi}{\partial x}= \frac{b \alpha}{2}  (2x+K^2y)\left(s^2+x^2+\frac{y^2}{q^2}+K^2xy  \right)^{\frac{\alpha}{2}-1}
\end{equation}
\begin{equation}
    \alpha_y=\frac{\partial \Psi}{\partial x}= \frac{b \alpha}{2}  \left(\frac{2y}{q^2}+K^2x\right)\left(s^2+x^2+\frac{y^2}{q^2}+K^2xy  \right)^{\frac{\alpha}{2}-1}
\end{equation}
The second derivatives of the projected potential are given below,
\begin{align}
    \Psi_{xx}=&\frac{\partial^2 \Psi}{\partial x^2}= b \alpha  \left(s^2+x^2+\frac{y^2}{q^2}+K^2xy  \right)^{\frac{\alpha}{2}-1} \nonumber \\ &+\frac{b \alpha}{2}\left(\frac{\alpha}{2}-1 \right)  \left(2x+K^2y\right)^2\left(s^2+x^2+\frac{y^2}{q^2}+K^2xy  \right)^{\frac{\alpha}{2}-2}
\end{align}
\begin{align}
    \Psi_{yy}=&\frac{\partial^2 \Psi}{\partial y^2}= \frac{b \alpha }{q^2} \left(s^2+x^2+\frac{y^2}{q^2}+K^2xy  \right)^{\frac{\alpha}{2}-1} \nonumber \\ &+\frac{b \alpha}{2}\left(\frac{\alpha}{2}-1 \right)  \left(\frac{2y}{q^2}+K^2x\right)^2\left(s^2+x^2+\frac{y^2}{q^2}+K^2xy  \right)^{\frac{\alpha}{2}-2}
\end{align}
\begin{align}
    \Psi_{xy}=& \Psi_{yx}=\frac{\partial^2 \Psi}{\partial x \partial y}=\frac{\partial^2 \Psi}{\partial x \partial y}= \frac{K^2 b \alpha }{2} \left(s^2+x^2+\frac{y^2}{q^2}+K^2xy  \right)^{\frac{\alpha}{2}-1} \nonumber \\ &+\frac{b \alpha}{2}\left(\frac{\alpha}{2}-1 \right)  \left(2x+K^2y\right)\left(\frac{2y}{q^2}+K^2x\right) \nonumber \\ & \times \left(s^2+x^2+\frac{y^2}{q^2}+K^2xy  \right)^{\frac{\alpha}{2}-2}
\end{align}

\section{Plummer Mass Distribution}
\label{appendix:plummers}
The gravitational potential of a Plummer sphere \citep{Plummer1911} is given by,
\begin{equation}
    \Phi(r)=-\frac{GM}{(r^2+a_P^2)^{1/2}}
\end{equation}
where $a_P$ is the characteristic width of the mass distribution.
 The projected lens potential for the above potential is,
\begin{equation}
    \Psi(\Vec{\theta})=\frac{D_{ds}}{D_s}\frac{2GM}{c^2 D_d}\ln(\theta^2+\theta_P^2)
\end{equation}
where $\theta_P={a_P}/{D_d}$ is the characteristic angular width.


\bsp    
\label{lastpage}
\end{document}